\documentclass[12pt]{iopart}
\newcommand{\lesssim}{\raisebox{0.3mm}{\em $\, <$} \hspace{-2.8mm}
\raisebox{-1.3mm}{\em $\sim \,$}}
\newcommand{\gtrsim}{\raisebox{0.3mm}{\em $\, >$} \hspace{-2.8mm}
\raisebox{-1.3mm}{\em $\sim \,$}}
\usepackage{times}
\usepackage{graphicx}

\begin{document}

\title{Neutrino factories --- Physics potential and present status }
\author{Osamu Yasuda}
\address{Department of Physics,
Tokyo Metropolitan University \\
Minami-Osawa, Hachioji, Tokyo 192-0397, Japan
\\E-mail: yasuda@phys.metro-u.ac.jp}
\begin{abstract}
I briefly review the recent status of research on physics potential
of neutrino factories including the discussions on parameter degeneracy.
\end{abstract}

\section{Introduction}

\begin{figure}
\hglue 2.5cm
\includegraphics[width=7cm]{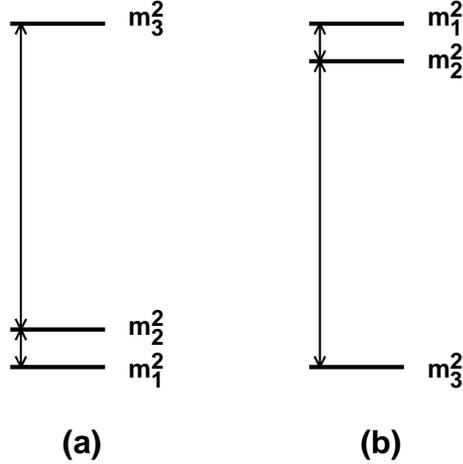}
\vglue 2.5cm 
\caption{Two mass patterns.\\
(a), (b) correspond to $\Delta m^2_{32}>0$, $\Delta m^2_{32}<0$,
respectively.}
\label{fig:pattern}
\end{figure}

The observation of atmospheric neutrinos (See, e.g.,
Ref.\cite{Kajita:2001mr})
and solar neutrinos (See, e.g., Ref.\cite{Bahcall:2000kh,Ahmad:2002ka})
gives the information on the mass squared differences
and the mixings, which can be written in the three flavor
framework of neutrino oscillations as
($|\Delta m_{32}^2|$, $\theta_{23}$) and
($\Delta m_{21}^2$, $\theta_{12}$), where I have adopted
the standard parametrization \cite{Hagiwara:pw}
for the $3\times3$ MNSP \cite{Maki:1962mu,Pontecorvo:1968fh} matrix.
On the other hand, the CHOOZ 
result \cite{Apollonio:1999ae} tells us that
$|\theta_{13}|$ has to be small ($\sin^22\theta_{13}\lesssim 0.1$).
So the MNSP matrix looks like

\begin{eqnarray}
U_{MNSP} \simeq
\left(
\begin{array}{ccc}
c_\odot & s_\odot &  \epsilon\\
-s_\odot/\sqrt{2} &
c_\odot/\sqrt{2} & 1/\sqrt{2}\\
s_\odot/\sqrt{2} &
-c_\odot/\sqrt{2} & 1/\sqrt{2}\\
\end{array}
\right),\nonumber
\end{eqnarray}
where I have used $\theta_{23}\simeq \pi/4$,
$\sin^22\theta_{12}\equiv\sin^22\theta_\odot\simeq 0.8$
and $|\epsilon|\ll 1$.
With the mass hierarchy $|\Delta m_{21}^2|\ll|\Delta m_{32}^2|$
there are two possible
mass patterns which are depicted in Fig. \ref{fig:pattern},
depending on whether $\Delta m_{32}^2$ is positive or negative.

The next thing one has to do in neutrino oscillation study
is to determine $\theta_{13}$,
the sign of $\Delta m_{32}^2$ and the CP phase $\delta$.
Among others, measurements of CP violation
\cite{Cabibbo:1977nk,Barger:1980jm,Pakvasa:1980bz,Bilenky:1981hf}
is the final goal of the neutrino oscillation experiments.
During the past few years a lot of research have been done
on the possibilities of future long baseline experiments.
One is a super-beam
experiment \footnote{The most realistic super beam project
for the moment is the JHF experiment \cite{Itow:2001ee}.}
and the other one is
a neutrino factory \cite{Geer:1997iz}. \footnote{For research
on a neutrino factory in the past, see, e.g.,
\cite{DeRujula:1998hd,Ayres:1999ug,Albright:2000xi,Adams:2001tv,
Hernandez:ki,Yasuda:2001ip}
and references therein.}
The former is super intense conventional
$\nu_\mu$ (or $\bar{\nu}_\mu$)
beam which is obtained from pion decays where
$\nu_e$ created in oscillation $\nu_\mu\rightarrow\nu_e$
is measured.  The latter is beam from muon decays
in a storage ring
and it consists of $\nu_\mu$ and $\bar{\nu}_e$
($\bar{\nu}_\mu$ and $\nu_e$) in the case of $\mu^-$ ($\mu^+$) decays.
To measure the oscillation probability
$P(E_\nu)\equiv P(\nu_e\rightarrow\nu_\mu)$ or
$P(E_{\bar\nu})\equiv P({\bar\nu}_e\rightarrow{\bar\nu}_\mu)$
at a neutrino factory, it is very important that
``wrong sign muons'' which are produced
in oscillation
$\bar\nu_e\rightarrow\bar\nu_\mu\rightarrow\mu^+$
(or $\nu_e\rightarrow\nu_\mu\rightarrow\mu^-$)
can be distinguished from right sign muons
which are produced from $\nu_\mu$ (or $\bar{\nu}_\mu$)
without oscillation.
The background fraction $f_B$ in the case of
super-beams is of order $10^{-2}$ \cite{Itow:2001ee},
while in the case of a neutrino factory $f_B$ is of order
$10^{-5}$ \cite{Cervera:2000vy}.  The advantage of a neutrino factory
is such low background fraction and neutrino factories are
expected to enable us to determine $\theta_{13}$ and
the sign of $\Delta m_{32}^2$ ($\delta$) for
$\sin^22\theta_{13}\gtrsim 10^{-5}$ ($10^{-3}$), respectively.

In this talk I will discuss measurements of
these quantities at neutrino factories.
The numbers $N_{\mbox{\rm wrong}}(\mu^\pm)$ of the wrong sign muons
are given by \cite{Geer:1997iz}
\begin{eqnarray}
N_{\mbox{\rm wrong}}(\mu^-)&=&
n_T{12E_\mu^2 \over \pi L^2 m_\mu^2}
\int d\left({E_\nu \over E_\mu}\right)
\left({E_\nu \over E_\mu}\right)^2
\left(1-{E_\nu \over E_\mu}\right)
\sigma_{\nu N}(E_\nu)P(E_\nu)\nonumber\\
N_{\mbox{\rm wrong}}(\mu^+)&=&
n_T{12E_\mu^2 \over \pi L^2 m_\mu^2}
\int d\left({E_{\bar\nu} \over E_\mu}\right)
\left({E_{\bar\nu} \over E_\mu}\right)^2
\left(1-{E_{\bar\nu} \over E_\mu}\right)
\sigma_{{\bar\nu} N}(E_{\bar\nu})P(E_{\bar\nu}),\nonumber
\end{eqnarray}
where $E_\mu$ is the muon energy, $L$ is the length of the neutrino path,
$n_T$ is the number of the target nucleons,
$\sigma_{\nu N}(E_\nu)$ and
$\sigma_{{\bar\nu} N}(E_{\bar\nu})$ are
the (anti-)neutrino nucleon cross sections given by
\begin{eqnarray}
\sigma_{\nu N}(E_\nu)&=&
\left({E_\nu \over \mbox{\rm GeV}}\right)
\times 0.67 \times 10^{-38}\mbox{\rm cm}^2\nonumber\\
\sigma_{{\bar\nu} N}(E_{\bar\nu})&=&
\left({E_{\bar\nu} \over \mbox{\rm GeV}}\right)
\times 0.33 \times 10^{-38}\mbox{\rm cm}^2,\nonumber
\end{eqnarray}
and $P(E_\nu)$ and $P(E_{\bar\nu})$ are the oscillation probabilities
$P(E_\nu)\equiv P(\nu_e\rightarrow\nu_\mu)$ and
$P(E_{\bar\nu})\equiv P({\bar\nu}_e\rightarrow{\bar\nu}_\mu)$.

\section{The sign of $\Delta m_{32}^2$}
\label{sec:sign}
As was mentioned in the Introduction, the mass pattern corresponds to
either Fig. \ref{fig:pattern}(a) or \ref{fig:pattern}(b),
depending on whether $\Delta m_{32}^2$ is
positive or negative.
Determination of this mass pattern is
important, since Figs. \ref{fig:pattern}(a) and \ref{fig:pattern}(b)
correspond to one and two mass states,
assuming that the lowest mass is almost zero. \footnote{The mixed dark
scenario in which neutrinos have masses of order 1 eV seems to be
disfavored by cosmology \cite{Fukugita:1999as,Elgaroy:2002bi}.
It has been reported that signals of neutrinoless double $\beta$ decay
was found \cite{Klapdor-Kleingrothaus:2001ke}.
If the claim is correct, then there must be overall shift to
the mass, since the mass range 0.05 eV -- 0.84 eV is larger
than $\sqrt{\Delta m^2_{\mbox{\rm\footnotesize atm}}}$.}

In the limit $\Delta m^2_{21}\rightarrow 0$
the oscillation probability essentially becomes
that of two flavors and is given by (See, e.g., \cite{Yasuda:1998sf})
\begin{eqnarray}
\hspace*{-20mm}
\left\{ \begin{array}{c}
P(\nu_e\rightarrow\nu_\mu)\\
P({\bar\nu}_e\rightarrow{\bar\nu}_\mu)\end{array}\right\}
=\sin^2\theta_{23}\sin^22\tilde{\theta}_{13}^{{}^{(\mp)}}
\sin^2\left({\Delta \tilde{E}_{32}^{{}^{(\mp)}}L \over 2}\right),
\label{eqn:prob0}
\end{eqnarray}
where $A\equiv\sqrt{2} G_F N_e$ stands for the matter effect
\cite{Mikheev:wj,Wolfenstein:1977ue} of the Earth,
$\tilde{\theta}_{13}^{{}^{(\pm)}}$ is the effective mixing angle in
matter given by
\begin{eqnarray}
\tan2\tilde{\theta}_{13}^{{}^{(\pm)}}\equiv {\Delta E_{32}\sin2\theta_{13}
\over \Delta E_{32}\cos2\theta_{13}\pm A},\nonumber
\end{eqnarray}
\noindent
\begin{eqnarray}
\Delta \tilde{E}_{32}^{{}^(\pm)}\equiv
\sqrt{\left(\Delta E_{32}\cos2\theta_{13}\pm A\right)^2
+\left(\Delta E_{32}\sin2\theta_{13}\right)^2},\nonumber
\end{eqnarray}
and $\Delta E_{32}\equiv \Delta m^2_{32}/2E\equiv(m^2_3-m^2_2)/2E$.

As can be seen from (\ref{eqn:prob0}), if $\Delta m_{32}^2>0$
then the effective mixing angle
$\tilde{\theta}_{13}^{{}^{(-)}}$ is enhanced and
$P(\nu_e\rightarrow\nu_\mu)$ increases.
On the other hand, if $\Delta m_{32}^2<0$ then 
$\tilde{\theta}_{13}^{{}^{(+)}}$ is enhanced and 
$P({\bar\nu}_e\rightarrow{\bar\nu}_\mu)$ increases.
So, at neutrino factories where baseline is relatively large and
therefore the matter effect plays an important role,
the sign of $\Delta m_{32}^2$ may be determined by looking at
the difference between neutrino and anti-neutrino events
which should reflect the difference between 
$P(\nu_e\rightarrow\nu_\mu)$ and
$P({\bar\nu}_e\rightarrow{\bar\nu}_\mu)$.

\begin{figure}
\vglue -0.7cm \hglue -4.8cm
\includegraphics[width=9cm]{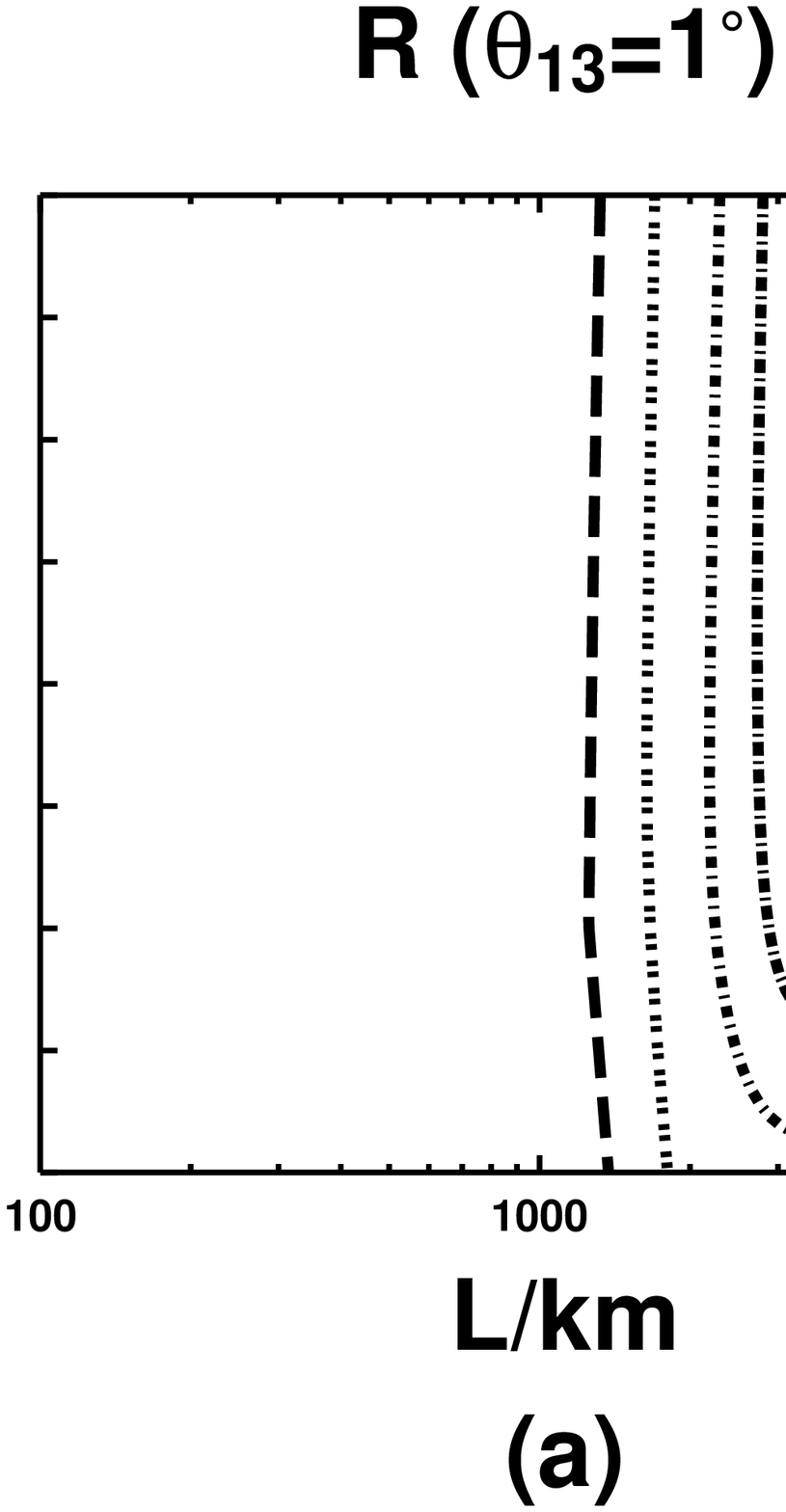}
\vglue -9.1cm \hglue 2.3cm
\includegraphics[width=9cm]{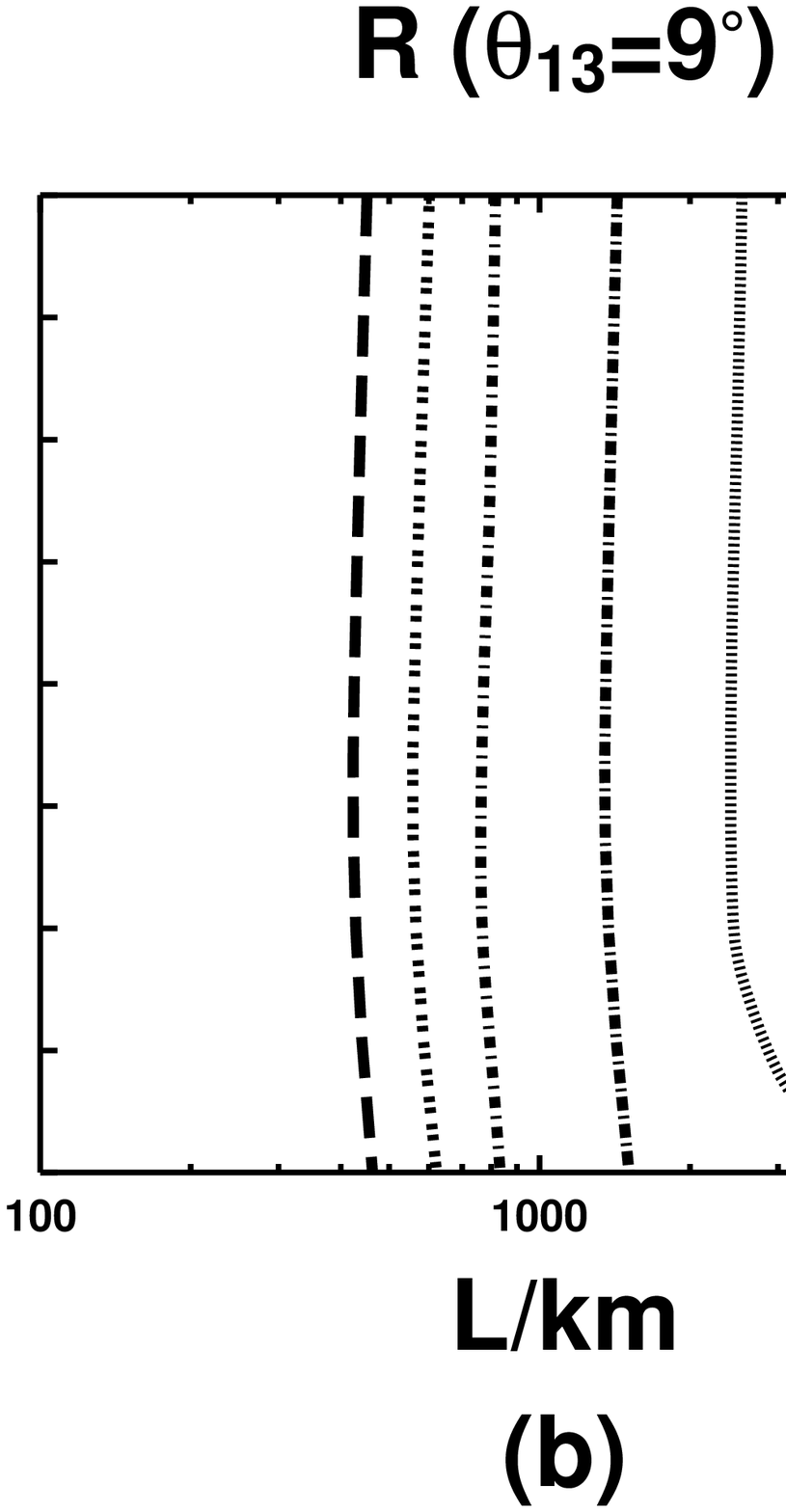}
\caption{The contour plot of the
ratio $R$ in (\ref{eqn:r}) to distinguish
sgn($\Delta m^2_{32}$) for
$\theta_{13}=1^\circ$, $\theta_{13}=9^\circ$, 
respectively \cite{Yasuda:1999zf}.}
\label{fig:fig2}
\end{figure}

Since the cross section $\sigma_{\nu N}$ and $\sigma_{{\bar\nu} N}$
are different (the ratio is 2 to 1), it is useful to look at the
quantity
\begin{eqnarray}
{N_\nu-2N_{\bar\nu} \over \delta(N_\nu-2N_{\bar\nu})}
={N_\nu-2N_{\bar\nu} \over \sqrt{N_\nu+4N_{\bar\nu}}}\nonumber
\end{eqnarray}
whose absolute value should be much larger than one
to demonstrate $\Delta m_{32}^2>0$ or $\Delta m_{32}^2<0$.
Now let me introduce the quantity
\begin{eqnarray}
R\equiv 
{\left[N_{\mbox{\rm wrong}}(\mu^-)-2N_{\mbox{\rm wrong}}(\mu^+)\right]^2
\over N_{\mbox{\rm wrong}}(\mu^-)+4N_{\mbox{\rm wrong}}(\mu^+)}.
\label{eqn:r}
\end{eqnarray}
If $R\gg 1$ then one can deduce the sign of
$\Delta m_{32}^2$.
The contour plot of $R$=const. is given in
Fig. \ref{fig:fig2}(a) and \ref{fig:fig2}(b)
for typical values of $\theta_{13}$ with
$\Delta m_{32}^2=3.5\times 10^{-3}$eV$^2>$0,
$\sin^22\theta_{23}=1.0$ \cite{Yasuda:1999zf}.
If $\sin^22\theta_{13}$ is not
smaller than $10^{-3}$, it is possible to determine the sign of
$\Delta m_{32}^2$.  Irrespective of the value of $\theta_{13}$,
$L\sim$ 5000km, $E_\mu=$ 50 GeV seem to optimize the signal,
as far as the quantity $R$ is concerned.
Similar results were obtained in \cite{Barger:2000yn,Barger:2000cp}
by considering the ratio
$N({\bar\nu}_e\rightarrow{\bar\nu}_\mu)
/N(\nu_e\rightarrow\nu_\mu)$.

\section{The magnitude of $\theta_{13}$}
\begin{figure}
\vglue 0.2cm \hglue 2.8cm
\includegraphics[width=9cm]{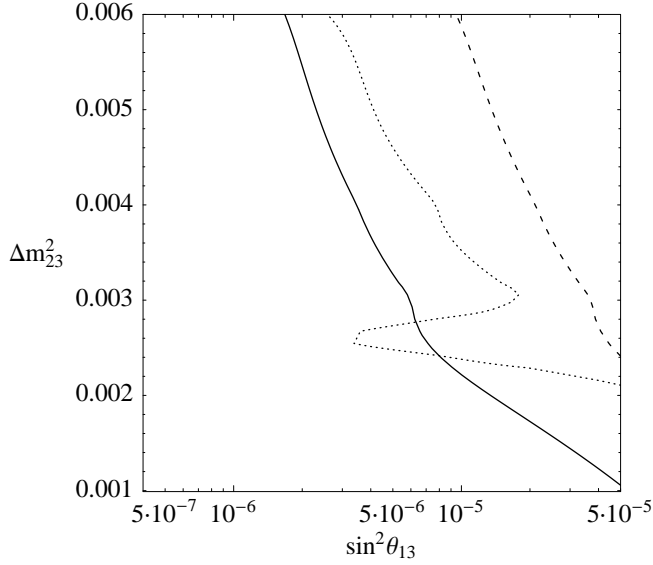}
\caption{Asymptotic sensitivity to $\sin^22\theta_{13}$ as a function of 
$\Delta m^2_{23}$ at 90\% CL for $L=732$ km (dashed lines),
3500 km (solid lines) and 7332 km (dotted lines), in the SMA-MSW solution. 
Background errors and detection efficiencies as well as
statistical errors are included \cite{Cervera:2000kp}.}
\label{fig:theta13}
\end{figure}

To establish non-zero value of $\theta_{13}$,
the following quantity has to be large:
\begin{eqnarray}
\Delta \chi^2\equiv
\sum_j
{\left[N_j(\nu_e\rightarrow\nu_\mu;\sin^22\theta_{13})
-N_j(\nu_e\rightarrow\nu_\mu;\sin^22\theta_{13}=0)\right]^2
\over N_j(\nu_e\rightarrow\nu_\mu;\sin^22\theta_{13})+
\left[f_B N_j(\nu_e\rightarrow\nu_\mu;\sin^22\theta_{13})\right]^2},
\nonumber
\end{eqnarray}
where $j$ stands for the label of energy bins and
$N_j$ stands for the numbers of events.
Taking realistic systematic errors and
detection efficiencies into consideration,
sensitivity of a neutrino factory to $\sin^22\theta_{13}$
was obtained in \cite{Cervera:2000kp} for baselines
$L$=732km, 3500km, 7332km with the muon energy $E_\mu$=50GeV,
the detector volume 40kt, useful muon decays $2\times10^{20}\mu$'s,
and is depicted in Fig. \ref{fig:theta13}.
Fig. \ref{fig:theta13} shows that neutrino factories
have sensitivity to $\sin^22\theta_{13}$ for
$\sin^22\theta_{13}\gtrsim 10^{-5}$.
Such high sensitivity is possible
because background fraction at neutrino factories
is small ($\sim10^{-5}$ \cite{Cervera:2000vy}).
The discussions in this section are devoted to
establishment of non-zero value of $\theta_{13}$,
and when it comes to determination of the precise
value of $\theta_{13}$, one has to take into account
the problem of parameter degeneracy, which will be
discussed later.

\section{Measurements of the CP phase $\delta$}
\subsection{Oscillation probabilities in vacuum and in matter}
To discuss the dependence of oscillation probability on $\delta$,
let me start with the oscillation probability in vacuum
which is given by
\begin{eqnarray}
\hspace*{-20mm}
\left\{ \begin{array}{c}
P(\nu_\alpha\rightarrow\nu_\beta)\\
P({\bar\nu}_\alpha\rightarrow{\bar\nu}_\beta)\end{array}\right\}
&=&\delta_{\alpha\beta}
-4\sum_{j<k}\mbox{\rm Re}\left(U_{\alpha j}U_{\beta j}^\ast
U_{\alpha k}^\ast U_{\beta k}\right)\sin^2\left(
{\Delta E_{jk}L \over 2}\right)\nonumber\\
&{\ }&\pm2\sum_{j<k}\mbox{\rm Im}\left(U_{\alpha j}U_{\beta j}^\ast
U_{\alpha k}^\ast U_{\beta k}\right)\sin\left(
\Delta E_{jk}L\right).\qquad\qquad\qquad
\label{eqn:prob}
\end{eqnarray}
%
The CP violation in vacuum is given by
\begin{eqnarray}
&{\ }&P(\nu_\alpha\rightarrow\nu_\beta)
-P({\bar\nu}_\alpha\rightarrow{\bar\nu}_\beta)
\nonumber\\
&=&4\sum_{j<k}\mbox{\rm Im}\left(U_{\alpha j}U_{\beta j}^\ast
U_{\alpha k}^\ast U_{\beta k}\right)\sin\left(
\Delta E_{jk}L\right)\nonumber\\
&=&4~J
\left[\sin\left(\Delta E_{12}L\right)
+\sin\left(\Delta E_{23}L\right)
+\sin\left(\Delta E_{31}L\right)
\right]\nonumber\\
&=&-16~J\sin\left(
{\Delta E_{31}L \over 2}\right)
\sin\left(
{\Delta E_{32}L \over 2}\right)
\sin\left(
{\Delta E_{21}L \over 2}\right)
,\nonumber
\end{eqnarray}
where
\begin{eqnarray}
J\equiv \mbox{\rm Im}\left(U_{\alpha 1}U_{\beta 1}^\ast
U_{\alpha 2}^\ast U_{\beta 2}\right)
\nonumber
\end{eqnarray}
is the Jarlskog factor, and
\begin{eqnarray}
\mbox{\rm Im}\left(U_{\alpha 1}U_{\beta 1}^\ast
U_{\alpha 2}^\ast U_{\beta 2}\right)=
\mbox{\rm Im}\left(U_{\alpha 2}U_{\beta 2}^\ast
U_{\alpha 3}^\ast U_{\beta 3}\right)=
\mbox{\rm Im}\left(U_{\alpha 3}U_{\beta 3}^\ast
U_{\alpha 1}^\ast U_{\beta 1}\right),\nonumber
\end{eqnarray}
which follows from the unitarity relations, has been used.
This Jarlskog factor, which is written as
\begin{eqnarray}
J=c_{13}\sin2\theta_{12}\sin2\theta_{13}\sin2\theta_{23}\sin\delta
\nonumber
\end{eqnarray}
in the standard parametrization \cite{Hagiwara:pw},
contains a small factor $\sin2\theta_{13}$ which
is constrained by the CHOOZ data ($\lesssim \sqrt{0.1}$).
So the Jarlskog factor is expected to be small.
In vacuum the asymmetry factor
\begin{eqnarray}
{\cal A}\equiv {P(\nu_\alpha\rightarrow\nu_\beta)
-P({\bar\nu}_\alpha\rightarrow{\bar\nu}_\beta) \over
P(\nu_\alpha\rightarrow\nu_\beta)
+P({\bar\nu}_\alpha\rightarrow{\bar\nu}_\beta)}
\label{eqn:asymmetry}
\end{eqnarray}
is a useful quantity to measure the CP phase $\delta$.
In vacuum the CP violation happens to be the same as the T violation
$P(\nu_\alpha\rightarrow\nu_\beta)-P(\nu_\beta\rightarrow\nu_\alpha)$.

On the other hand, in the presence of matter, the expression
(\ref{eqn:prob}) for the probability is modified. 
The eigen matrix in matter
can be formally diagonalized by a unitary matrix $\tilde{U}^{{}^{(\mp)}}$:
\begin{eqnarray}
\hspace*{-20mm}
U^{{}^{(\ast)}}\,\mbox{\rm diag} \left(E_1,E_2,E_3 \right) U^{{}^{(\ast)}-1}
\pm\mbox{\rm diag} \left(A,0,0 \right)
=\tilde{U}^{{}^{(\mp)}}\,\mbox{\rm diag} \left(\tilde{E}^{{}^{(\mp)}}_1,
\tilde{E}^{{}^{(\mp)}}_2,\tilde{E}^{{}^{(\mp)}}_3\right)
\tilde{U}^{{}^{(\mp)-1}},
\nonumber
\end{eqnarray}
where the sign for the matter term $A\equiv \sqrt{2}G_F N_e$
is reversed and the complex conjugate $U^{{}^{\ast}}$
of the unitary matrix is used instead of $U$
for antineutrinos ${\bar\nu}_\alpha$, so the
unitary matrix $\tilde{U}^{{}^{(+)}}$ and the eigenvalues
${\tilde{E}^{{}^{(+)}}}_j$ for ${\bar\nu}_\alpha$ are different
from $\tilde{U}^{{}^{(-)}}$ and $\tilde{E}^{{}^{(-)}}_j$ for
neutrinos $\nu_\alpha$.  Assuming constant density,
the oscillation probability can be written as
\begin{eqnarray}
\hspace*{-20mm}
\left\{ \begin{array}{c}
P(\nu_\alpha\rightarrow\nu_\beta)\\
P({\bar\nu}_\alpha\rightarrow{\bar\nu}_\beta)\end{array}\right\}
&=&\delta_{\alpha\beta}
-4\sum_{j<k}\mbox{\rm Re}\left(\tilde{U}^{{}^{(\mp)}}_{\alpha j}\tilde{U}^{{}^{(\mp)\ast}}_{\beta j}
\tilde{U}^{{}^{(\mp)\ast}}_{\alpha k} \tilde{U}^{{}^{(\mp)}}_{\beta k}\right)\sin^2\left(
{\Delta \tilde{E}^{{}^{(\mp)}}_{jk}L \over 2}\right)\nonumber\\
&{\ }&+2\sum_{j<k}\mbox{\rm Im}\left(\tilde{U}^{{}^{(\mp)}}_{\alpha j}\tilde{U}^{{}^{(\mp)\ast}}_{\beta j}
\tilde{U}^{{}^{(\mp)\ast}}_{\alpha k} \tilde{U}^{{}^{(\mp)}}_{\beta k}\right)\sin\left(
\Delta \tilde{E}^{{}^{(\mp)}}_{jk}L\right),
\label{eqn:prob2}
\end{eqnarray}
as in the case of the probability in vacuum, and
$\Delta{\tilde{E}^{{}^{(\mp)}}}_{jk}\equiv{\tilde{E}^{{}^{(\mp)}}}_j
-{\tilde{E}^{{}^{(\mp)}}}_k$.
Therefore
the asymmetry factor ${\cal A}$ in (\ref{eqn:asymmetry})
is not illuminating to
see the CP violation in matter, since it
contains both terms which vanish in the limit $\delta\rightarrow 0$
and those which do not:
\begin{eqnarray}
\hspace*{-25mm}
&{\ }&P(\nu_\alpha\rightarrow\nu_\beta)-
P({\bar\nu}_\alpha\rightarrow{\bar\nu}_\beta)\nonumber\\
\hspace*{-25mm}
&=&-4\sum_{j<k}\mbox{\rm Re}\left[\tilde{U}^{{}^{(-)}}_{\alpha j}\tilde{U}^{{}^{(-)\ast}}_{\beta j}
\tilde{U}^{{}^{(-)\ast}}_{\alpha k} \tilde{U}^{{}^{(-)}}_{\beta k}\sin^2\left(
{\Delta \tilde{E}^{{}^{(-)}}_{jk}L \over 2}\right)
-\tilde{U}^{{}^{(+)}}_{\alpha j}\tilde{U}^{{}^{(+)\ast}}_{\beta j}
\tilde{U}^{{}^{(+)\ast}}_{\alpha k} \tilde{U}^{{}^{(+)}}_{\beta k}\sin^2\left(
{\Delta {\tilde{E}^{{}^{(+)}}}_{jk}L \over 2}\right)\right]\nonumber\\
\hspace*{-25mm}
&{\ }&+2\sum_{j<k}\mbox{\rm Im}\left[\tilde{U}^{{}^{(-)}}_{\alpha j}\tilde{U}^{{}^{(-)\ast}}_{\beta j}
\tilde{U}^{{}^{(-)\ast}}_{\alpha k} \tilde{U}^{{}^{(-)}}_{\beta k}\sin\left(
\Delta \tilde{E}^{{}^{(-)}}_{jk}L\right)
-\tilde{U}^{{}^{(+)}}_{\alpha j}\tilde{U}^{{}^{(+)\ast}}_{\beta j}
\tilde{U}^{{}^{(+)\ast}}_{\alpha k} \tilde{U}^{{}^{(+)}}_{\beta k}\sin\left(
\Delta {\tilde{E}^{{}^{(+)}}}_{jk}L\right)\right].\nonumber
\end{eqnarray}
It has been pointed out
\cite{Kuo:1987km,Krastev:yu,Toshev:vz,Toshev:ku,Arafune:1996bt,
Sato:1997st, Koike:1997dh, Barger:1999hi, Koike:1999hf, Yasuda:1999uv,
Sato:1999wt, Koike:1999tb, Harrison:2000df, Sato:1999sb,
Yokomakura:2000sv, Koike:2001kv, Akhmedov:2001kd, Rubbia:2001pk,
Ota:2001cz, Guo:2001yt, Bueno:2001jd, Minakata:2002qi}
that T violation
is useful to probe the CP violating phase in the presence of matter.
In fact from (\ref{eqn:prob2}) I have T violation in matter:
\begin{eqnarray}
\Delta P&\equiv&
P(\nu_\alpha\rightarrow\nu_\beta)- P(\nu_\beta\rightarrow\nu_\alpha)
\nonumber\\
&=&4\sum_{j<k}\mbox{\rm Im}\left(\tilde{U}^{{}^{(-)}}_{\alpha j}\tilde{U}^{{}^{(-)\ast}}_{\beta j}
\tilde{U}^{{}^{(-)\ast}}_{\alpha k} \tilde{U}^{{}^{(-)}}_{\beta k}\right)\sin\left(
\Delta \tilde{E}^{{}^{(-)}}_{jk}L\right)\nonumber\\
&=&4~\tilde{J}^{{}^{\,\,(-)}}
\left[\sin\left(\Delta \tilde{E}^{{}^{(-)}}_{12}L\right)
+\sin\left(\Delta \tilde{E}^{{}^{(-)}}_{23}L\right)
+\sin\left(\Delta \tilde{E}^{{}^{(-)}}_{31}L\right)
\right]\nonumber\\
&=&-16~\tilde{J}^{{}^{\,\,(-)}}\sin\left(
{\Delta \tilde{E}^{{}^{(-)}}_{31}L \over 2}\right)
\sin\left(
{\Delta \tilde{E}^{{}^{(-)}}_{32}L \over 2}\right)
\sin\left(
{\Delta \tilde{E}^{{}^{(-)}}_{21}L \over 2}\right)
,\nonumber
\end{eqnarray}
where
\begin{eqnarray}
\tilde{J}^{{}^{\,\,(-)}}\equiv\mbox{\rm Im}\left(\tilde{U}^{{}^{(-)}}_{\alpha 1}\tilde{U}^{{}^{(-)\ast}}_{\beta 1}
\tilde{U}^{{}^{(-)\ast}}_{\alpha 2} \tilde{U}^{{}^{(-)}}_{\beta 2}\right)
\nonumber
\end{eqnarray}
is the modified Jarlskog factor in matter.
It is known \cite{Naumov:1991rh,Naumov:1991ju,Harrison:2000df} that
this modified Jarlskog factor can be written in terms of
$J$ as \footnote{A different form of the
quantity $\tilde{J}^{{}^{\,\,(\mp)}}/J$ was given
in \cite{Krastev:yu}.}
\begin{eqnarray}
\tilde{J}^{{}^{\,\,(\mp)}}=
{\Delta E_{31}\Delta E_{32}\Delta E_{21} \over
\Delta \tilde{E}^{{}^{(\mp)}}_{31}
\Delta \tilde{E}^{{}^{(\mp)}}_{32}
\Delta \tilde{E}^{{}^{(\mp)}}_{21}} ~J.
\end{eqnarray}
Unfortunately it is known that measurements of T violation
is experimentally difficult, and most of the discussions
in the past have been focused on a kind of indirect measurements
of the CP phase, i.e., one deduces the values of
$\delta$ etc. by comparing the energy spectra of the data and of
the theoretical prediction with neutrino oscillations
assuming the three flavor mixing.

\subsection{Establishing non-zero value of $\delta$}
\begin{figure}
\vglue -1.7cm \hglue -2.1cm
\includegraphics[width=9cm]{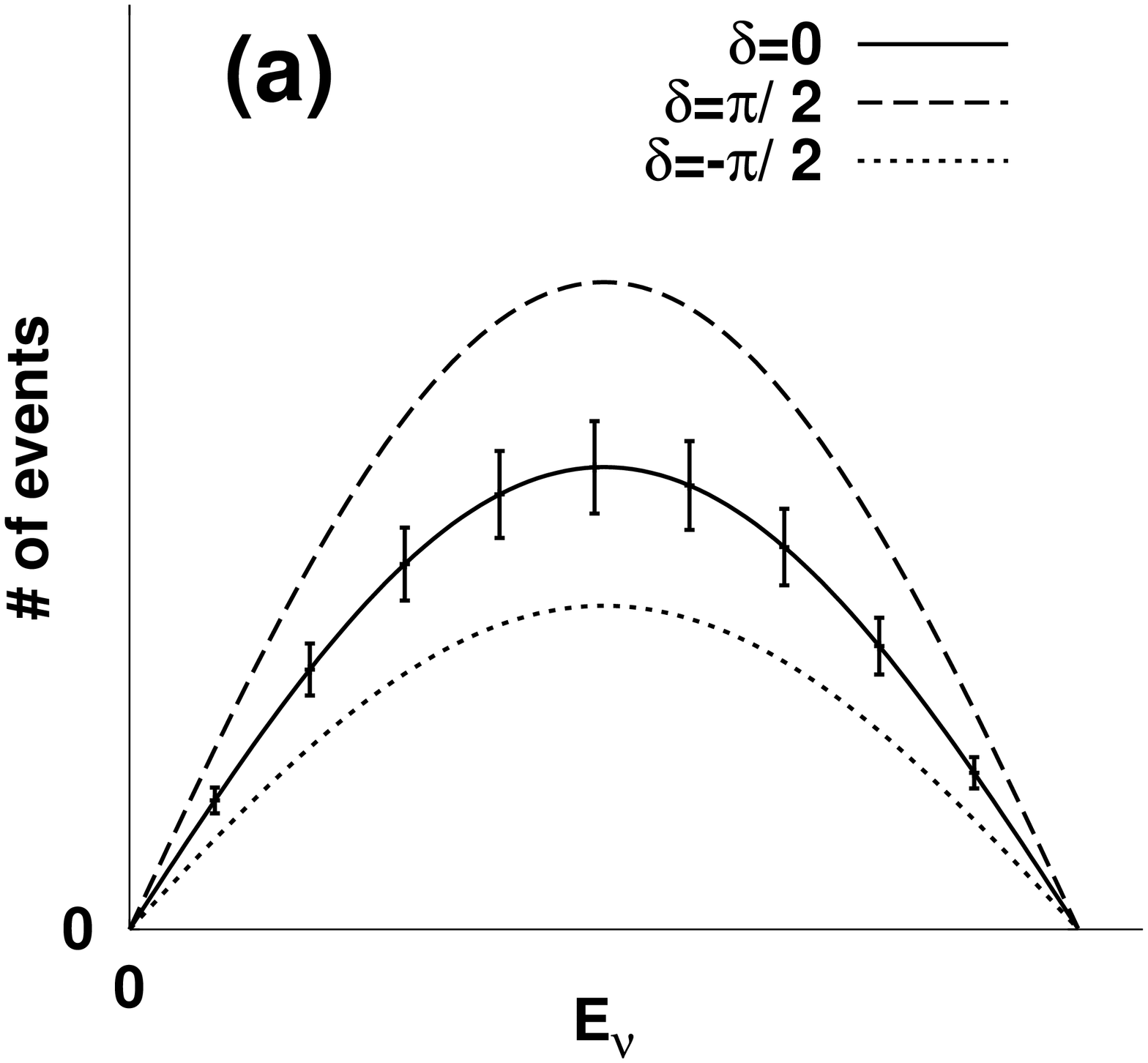}
\vglue -9.0cm \hglue 5.5cm
\includegraphics[width=9cm]{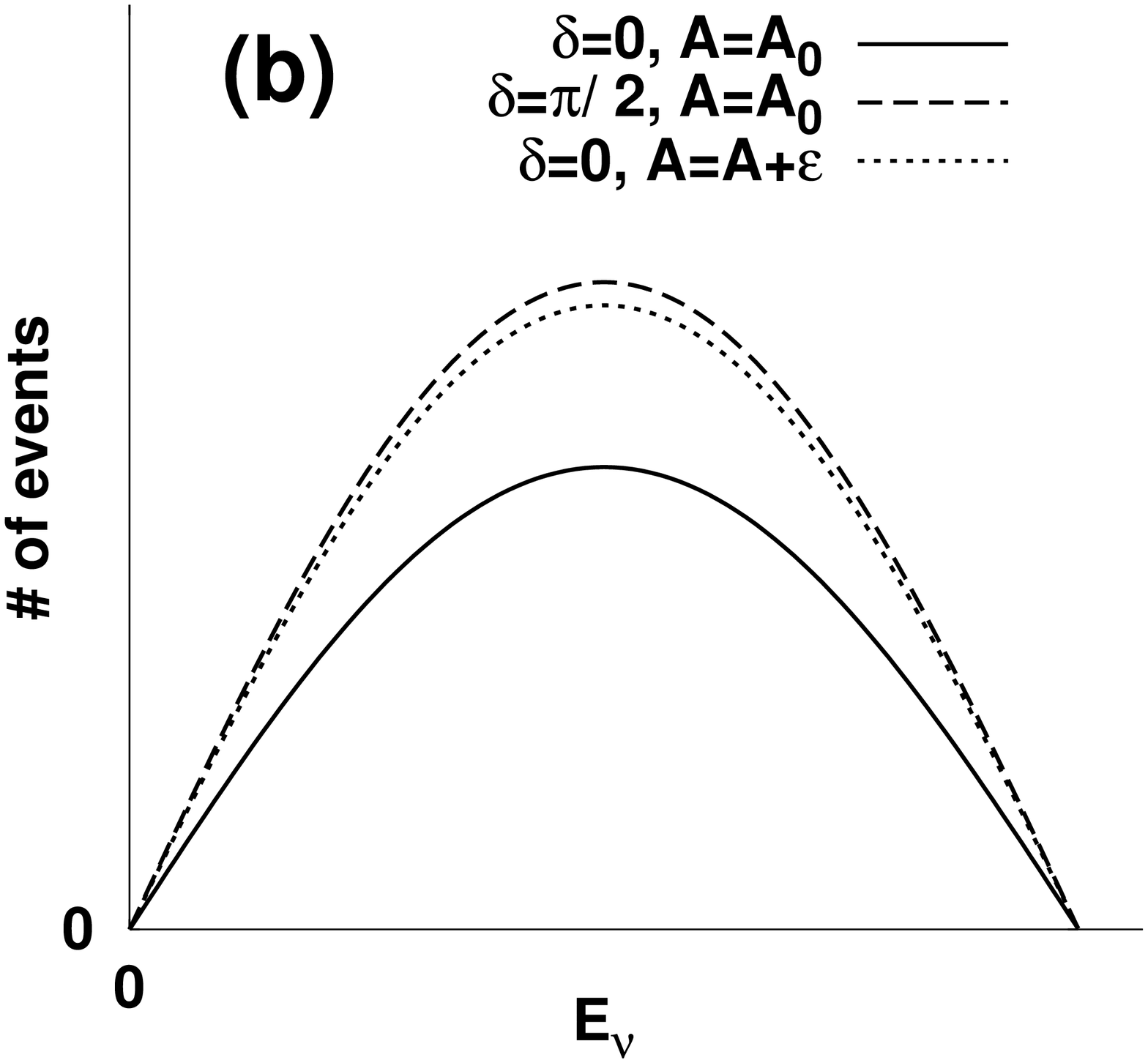}
\caption{(a) Indirect measurements of CP violation.
To show $\delta\ne 0$, the deviation of the numbers of events
with $\delta=\pm\pi/2$ has to be larger than the errors.
(b) Correlation of errors.
This is a hypothetical case where correlation between
$\delta$ and the matter effect $A\equiv \sqrt{2}G_F N_e$ is strong.
Even if the deviation of $\delta=\pi/2$ with $A=A_0$ (dashed lines)
from the case of $\delta=0$ with $A=A_0$ (solid lines) is large,
the number of events with $\delta=0$, $A=A_0+\epsilon$
(dotted lines) mimics the case of $\delta=\pi/2$ with $A=A_0$,
so in this case there is no way to show $\delta\ne 0$.}
\label{fig:corr}
\end{figure}

To establish non-zero value of $\delta$,
what one has to do is to show that the deviation
of the numbers of events with $\delta\ne 0$ from those with $\delta=0$
has to be larger than the errors
(cf. Fig. \ref{fig:corr}(a)).
Also, since there are other oscillation parameters as
well as the density (cf. \cite{Geller:2001ix}) of the Earth
whose values are not exactly known,
one has to take into account correlations of errors of these
parameters.\footnote{Correlations of errors at neutrino factories
were studied in
\cite{Cervera:2000kp,Bueno:2000fg,Koike:2002jf,Freund:2001ui,Pinney:2001xw}.}
To explain the situation, let me consider an example
of the correlation of the CP phase $\delta$ and the matter effect
$A\equiv \sqrt{2}G_F N_e$.  First of all, to be able to
show that $\delta\ne 0$,
the theoretical estimation for the numbers of events
with $\delta\ne 0$ and $\delta=0$ under an experimental setup has to
look like Fig. \ref{fig:corr}(a),
where the deviation of the numbers of events with $\delta=\pm\pi/2$
is larger than the error of that with $\delta=0$.
However, if the number of events with $\delta=0$ with $A=A_0+\epsilon$
(dotted lines) mimics the case of $\delta=\pi/2$ with $A=A_0$
as is depicted in Fig. \ref{fig:corr}(b),
then there is no way to distinguish the case of
$\delta=\pi/2$ with $A=A_0$ from that of $\delta=0$ with
$A=A_0+\epsilon$.
Therefore, one has to make sure that
the case with $\delta=\pi/2$ is distinct from that with
$\delta=0$, no matter which value the oscillation parameters
and the density of the Earth may take within the allowed region
of the parameters.  In other words, one has to choose
the baseline and the muon energy of a neutrino factory
such that the correlations of errors of the parameters are
small.

Thus I introduce the following quantity
to see the significance of the case with nonvanishing $\delta$:
\begin{eqnarray}
&&
\hspace*{-25mm}
\Delta\chi^2 \equiv
\min_{{\ }_{\overline{\theta_{k\ell}},
\overline{\Delta m^2_{k\ell}},\overline{A}}}
\displaystyle\sum_j \left\{
{\left[
N_j(\nu_e\rightarrow\nu_\mu)-\bar{N}_j(\nu_e\rightarrow\nu_\mu)
\right]^2 \over \sigma^2_j}
+{\left[N_j(\bar{\nu}_e\rightarrow\bar{\nu}_\mu)
-\bar{N}_j(\bar{\nu}_e\rightarrow\bar{\nu}_\mu)
\right]^2 \over \sigma^2_j}\right.\nonumber\\
&&\hspace*{-5mm}{}+\left.{\left[
N_j(\nu_\mu\rightarrow\nu_\mu)-\bar{N}_j(\nu_\mu\rightarrow\nu_\mu)
\right]^2 \over \sigma^2_j}
+ {\left[
N_j(\bar{\nu}_\mu\rightarrow\bar{\nu}_\mu)
-\bar{N}_j(\bar{\nu}_\mu\rightarrow\bar{\nu}_\mu)
\right]^2 \over \sigma^2_j}\right\},
\label{eqn:chi2}
\end{eqnarray}
where $N_j(\nu_\alpha\rightarrow\nu_\beta)
\equiv N_j(\nu_\alpha\rightarrow\nu_\beta;
\theta_{k\ell},\Delta m_{k\ell}^2,\delta,A)$,
$\bar{N}_j(\nu_\alpha\rightarrow\nu_\beta)
\equiv \bar{N}_j(\nu_\alpha\rightarrow\nu_\beta;
\overline{\theta_{k\ell}},\overline{\Delta m_{k\ell}^2},
\bar{\delta}=0,\bar{A})$ stand for the numbers of events
of the data and of the theoretical prediction with a vanishing
CP phase, respectively, and $\sigma^2_j$ stands for the error
which is given by the sum of the statistical and systematic
errors.  At neutrino factories appearance and disappearance
channels for $\nu$ and $\bar{\nu}$ are observed,
and I have included the numbers of events of all the
channels in (\ref{eqn:chi2}) to gain statistics.
In the present analysis, $N_j(\nu_\alpha\rightarrow\nu_\beta)$
is substituted by theoretical prediction with a CP phase $\delta$
and $\Delta\chi^2$ obviously vanishes if
$\delta=0$.\footnote{This is the reason why the quantity
in (\ref{eqn:chi2}) is denoted as $\Delta\chi^{2}$ instead of
absolute $\chi^2$.  $\Delta\chi^2$ represents
deviation from the best fit point rather than the goodness
of fit.}
The quantity $\Delta\chi^2$ reflects the strength of
the correlation of the parameters, i.e., if
$\Delta\chi^2$ turns out to be very small for
a certain value of $\bar{A}$ then the correlation
between $\delta$ and $A$ would be very strong and
in that case there would be no way to show
$\delta\ne0$.
To reject a hypothesis ``$\delta=0$'' at the 3$\sigma$
confidence level, I demand
\begin{eqnarray}
\Delta\chi^2\ge\Delta\chi^2(3\sigma CL)
\label{eqn:3sigma}
\end{eqnarray}
where the right hand side stands for the value of
$\chi^2$ which gives the probability 99.7\%
in the $\chi^2$ distribution with a certain degrees freedom,
and $\Delta\chi^2(3\sigma CL)$=20.1 for 6 degrees freedom.
From (\ref{eqn:3sigma}) I get the condition for the detector size
to reject a hypothesis ``$\delta=0$'' at 3$\sigma$CL.

We note in passing that indirect measurements
of CP violation are also considered in the $B^0-\bar{B}^0$
system.\footnote{The term "indirect" in the B system is different
from that in neutrino oscillations.}
To measure the phase $\phi_1$, direct CP violating process
$B(\bar{B})\rightarrow J/\psi\, K_{s}$ is
used \cite{Carter:1981tk,Bigi:1981qs},
while those to measure $\phi_2$ and $\phi_3$ are
$B\rightarrow 2\pi$ \cite{Gronau:1990ka} and
$B\rightarrow D K$ \cite{Gronau:1991dp}, which are
not necessarily CP--odd processes.
Their strategy is to start with the three flavor framework,
to use the most effective process, which may or may not be
CP--odd, to determine the CP phases and to check unitarity
or consistency of the three flavor hypothesis.

\begin{table}
\vglue 0.9cm
\hglue 0.5cm
\begin{tabular}{|p{8mm}|p{17mm}|p{10mm}|p{6mm}|p{23mm}|p{12mm}|p{14mm}|p{18mm}|}
\hline
{Ref.}&
{correlations of $\theta_{ij}$,$m_{ij}^2$}&
{$|\Delta A/A|$}&
{$f_B$}&
{$\Delta m_{21}^2/10^{-5}$eV$^2$}&
{$E_{th}$/GeV}&
{optimized $E_{\mu}$/GeV}&
{optimized $L$/km}\\
\hline
\hline
KOS \cite{Koike:2002jf}&
included&
10\%&
0&
5&
1&
$\lesssim$6&
600 -- 800\\
\cline{5-8}
&&&&10&1&
$\lesssim$50&
500 -- 2000\\
\hline
FHL \cite{Freund:2001ui}&
included&
0&
0&
10&
4&
30 -- 50&
2800 -- 4500\\
\hline
PY \cite{Pinney:2001xw}&
included&
5\%&
$10^{-5}$&
3.2&
0.1&
$\sim$50&
$\sim$3000\\
\cline{4-4}\cline{7-8}
&&&$10^{-3}$&
&&$\sim$20&
$\sim$1000\\
\cline{3-4}\cline{7-8}
&&10\%&
$10^{-5}$&
&&$\sim$15&
$\sim$1000\\
\cline{4-4}\cline{7-8}
&&&$10^{-3}$&
&&$\sim$10&
$\sim$800\\
\cline{3-4}\cline{7-8}
&&20\%&
$10^{-5}$&
&&$\sim$8&
$\sim$500\\
\cline{4-4}\cline{7-8}
&&&$10^{-3}$&
&&$\sim$6&
$\sim$500\\
\hline
H \cite{huber}&
included&
10\%&
0&
10&
0.1&
$\gtrsim$20&
$\gtrsim$2000\\
\cline{5-5}\cline{7-8}
&&&&3.5&&
$\sim$25&
$\sim$1500\\
\hline
\end{tabular}
\caption{Comparison of different works
The reference value is
$\sin^{2}2\theta_{13}=0.1$, $10^{21} \mu$s.}
\label{tbl:table1}
\end{table}


\begin{figure}
\vglue -1.3cm \hglue 1.0cm
\includegraphics[width=7cm]{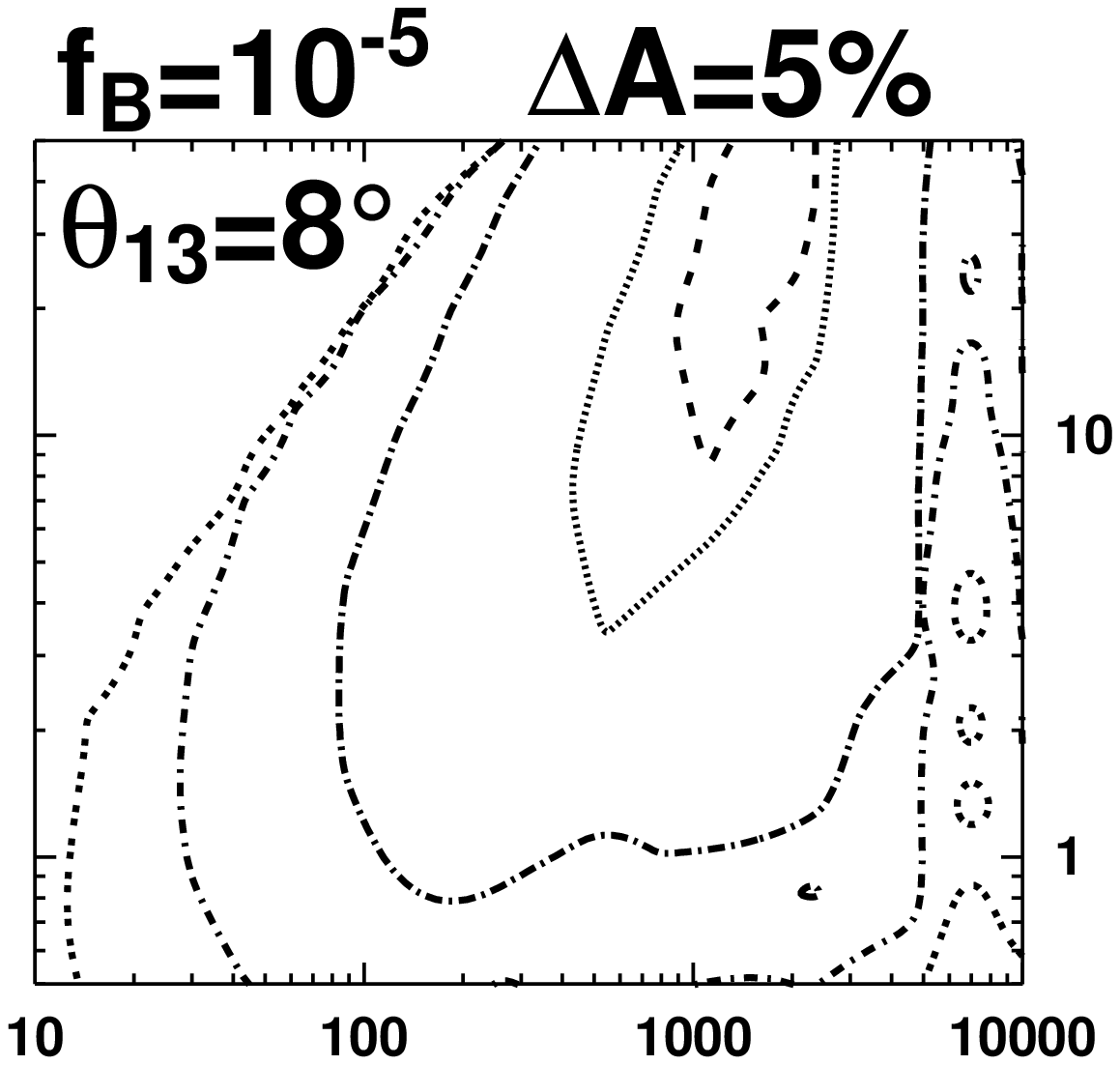}
\vglue -7.9cm \hglue 5.8cm
\includegraphics[width=7cm]{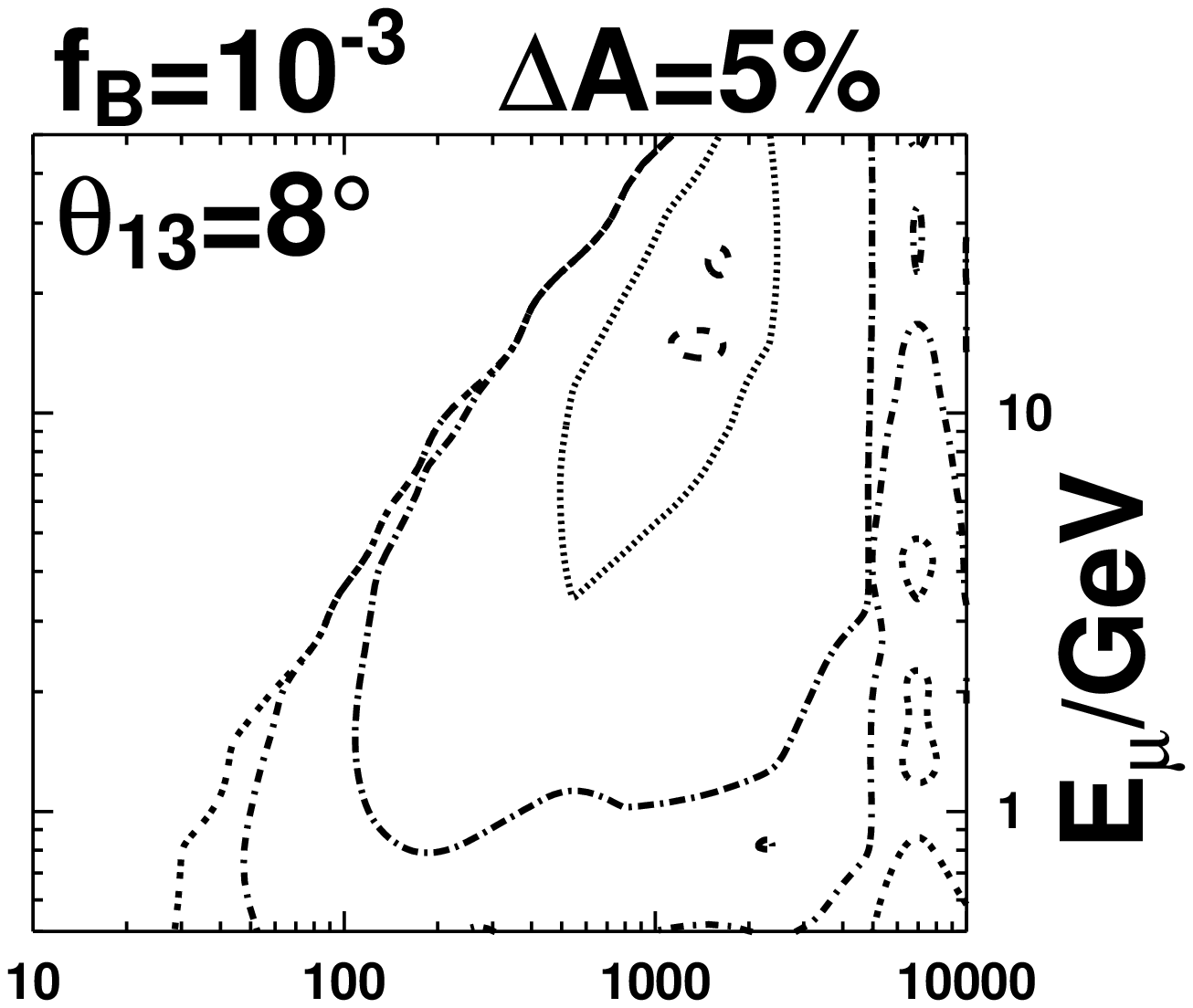}
\vglue -3.5cm \hglue 1.0cm
\includegraphics[width=7cm]{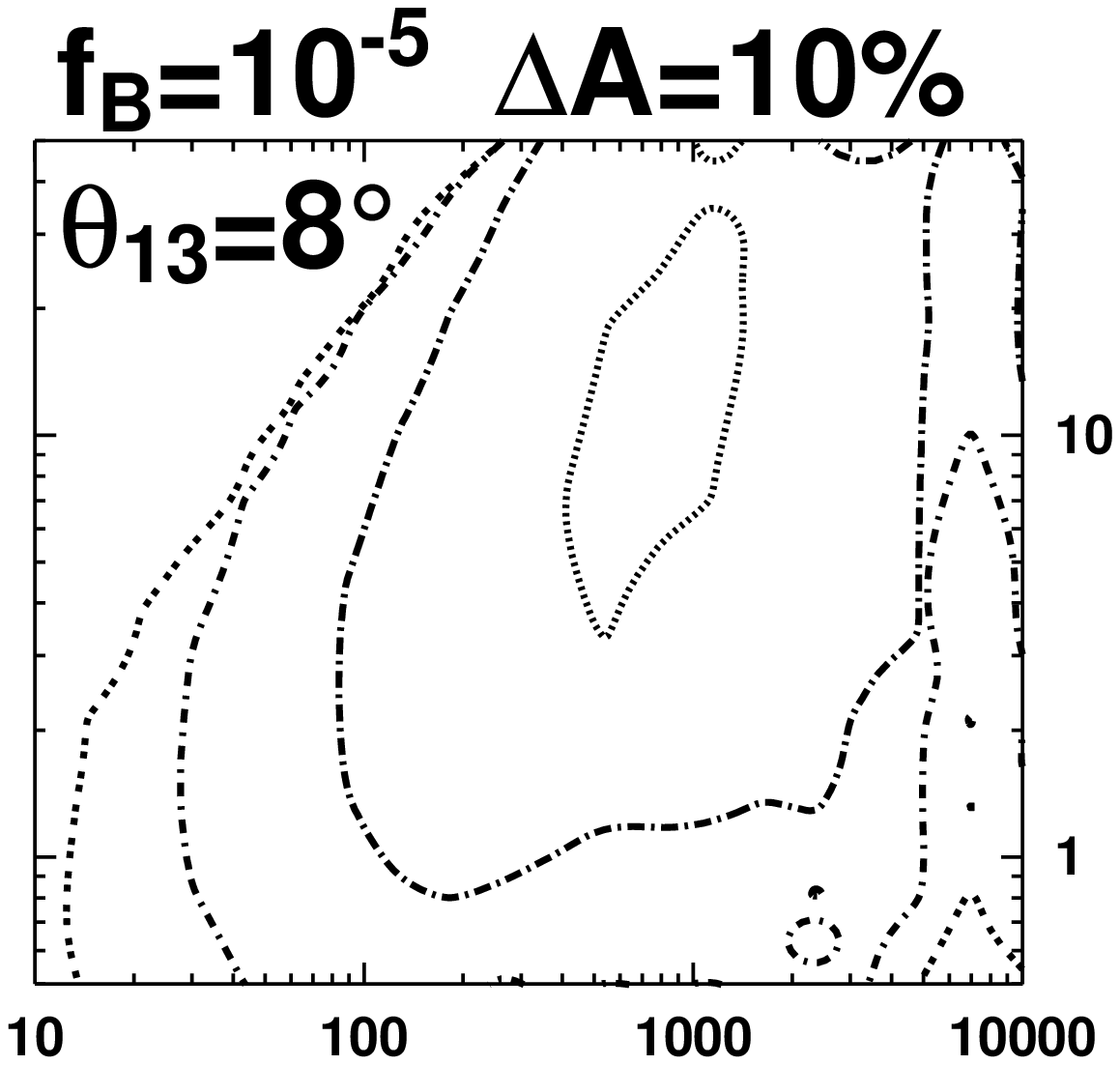}
\vglue -7.9cm \hglue 5.8cm
\includegraphics[width=7cm]{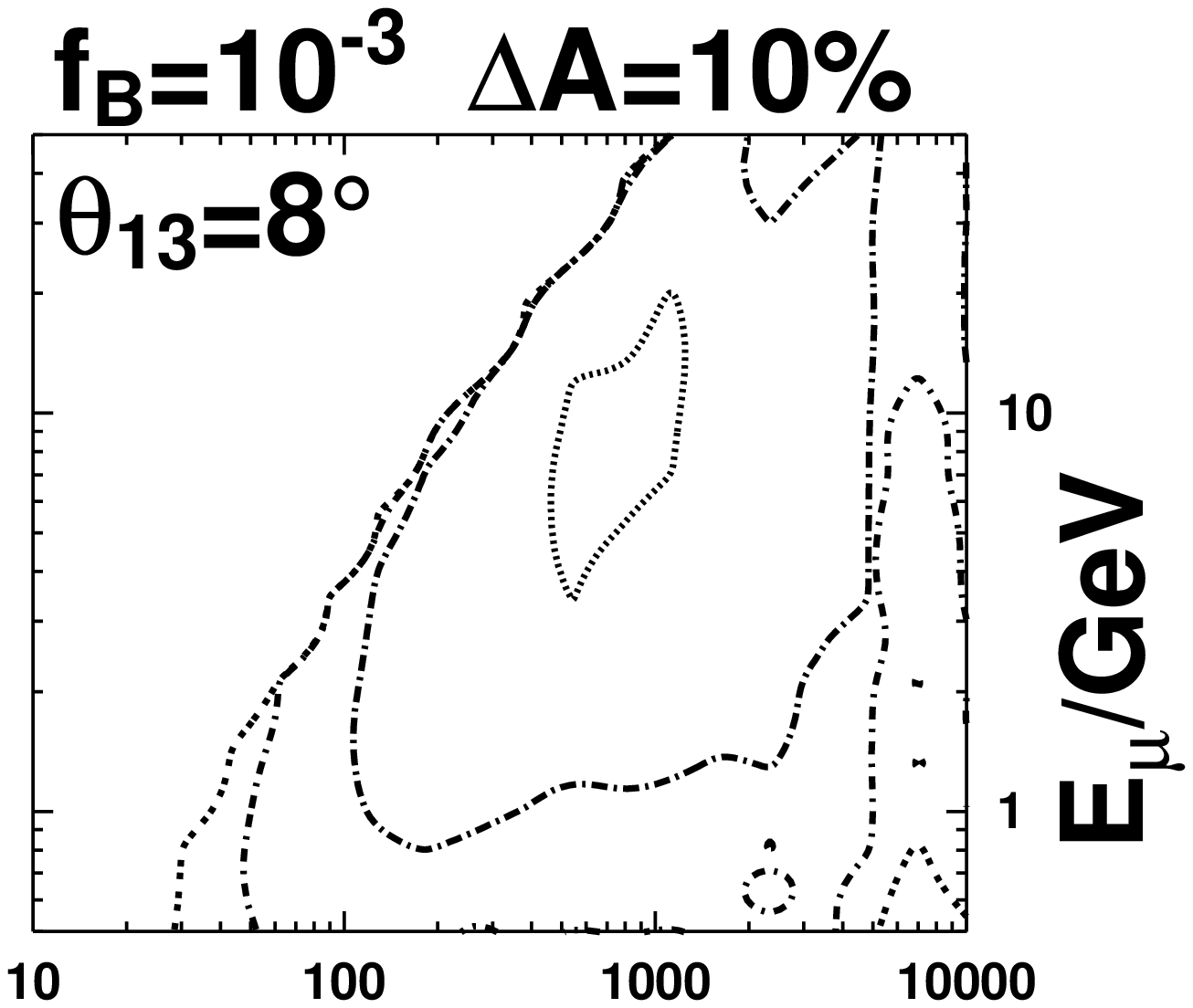}
\vglue -3.5cm \hglue 1.0cm
\includegraphics[width=7cm]{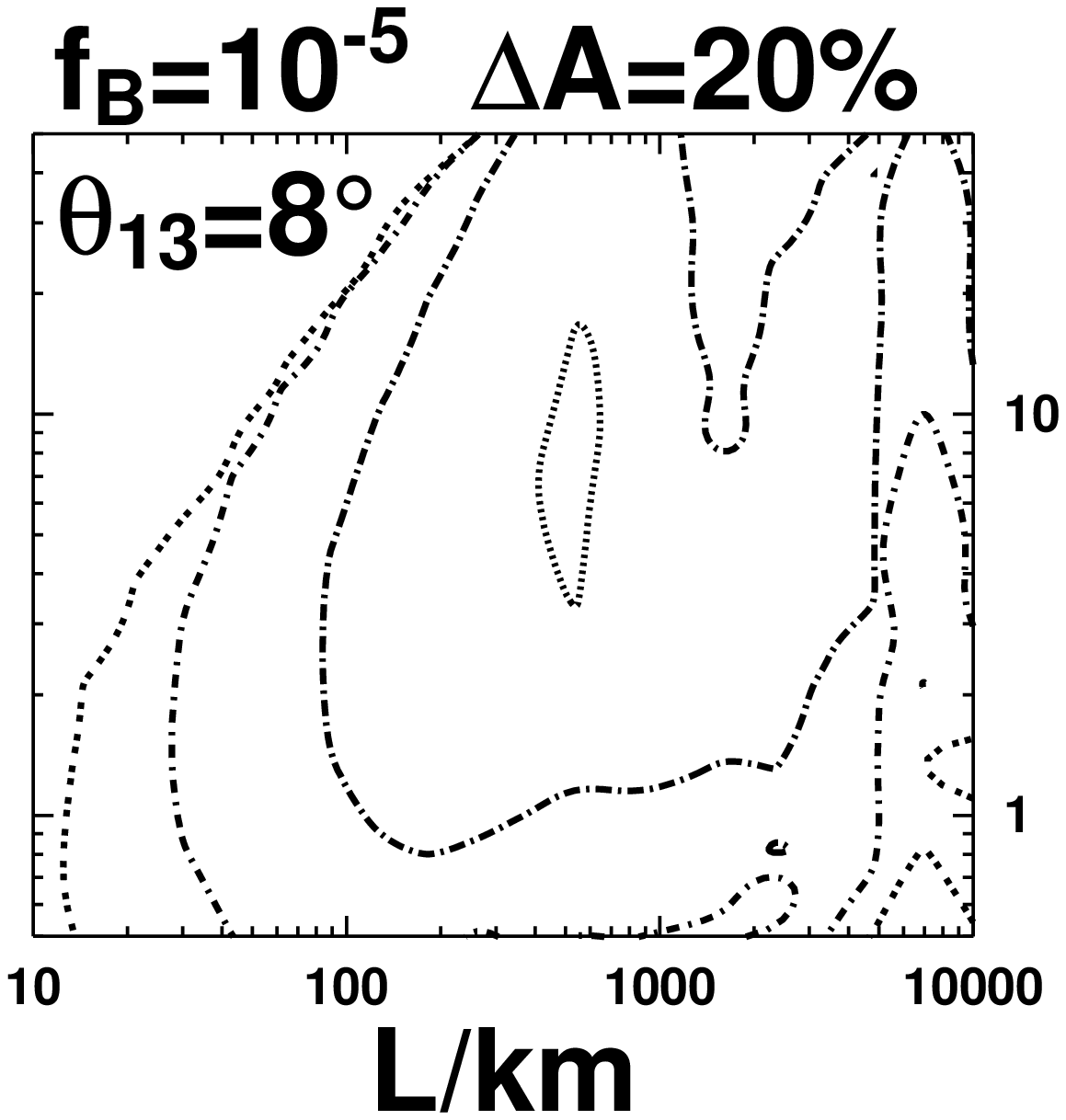}
\vglue -7.9cm \hglue 5.8cm
\includegraphics[width=7cm]{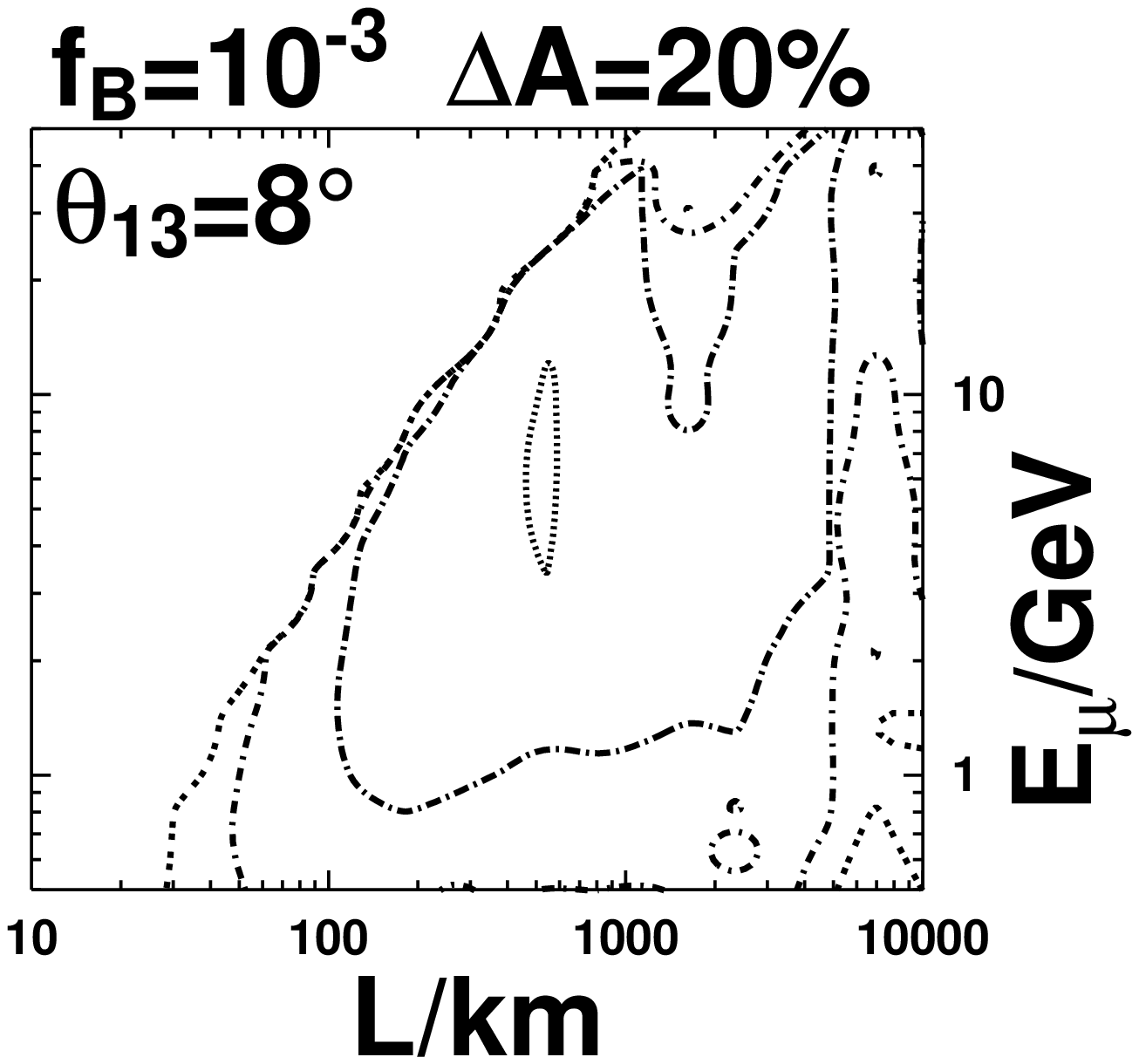}
\vglue -4.0cm\hglue -7.8cm
\includegraphics[width=14cm]{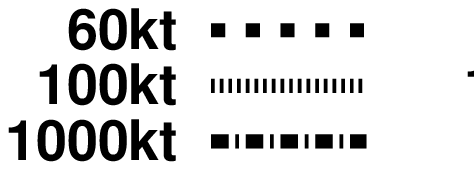}
\vglue -12.0cm \hglue 3.3cm
\caption{\small
The contour plot of equi-number of data size required
to reject a hypothesis $\bar{\delta}=0$ at $3\sigma$
with
the background fraction $f_B=10^{-5}$ or $10^{-3}$
and the uncertainty of the matter effect $\Delta A$=5\%, 10\% or 20\%.
\cite{Pinney:2001xw}}
\label{fig:Fig2}
\vglue -0.5cm 
\end{figure}

The optimized muon energy $E_\mu$ and baseline $L$
for the measurement of CP the phase at a neutrino
factory have been investigated by several groups
by taking into consideration the correlations of
$\delta$ and all other parameters and the results
are summarized in Table \ref{tbl:table1}.
The results in Ref.\cite{Pinney:2001xw} are given in Fig. \ref{fig:Fig2},
which shows that the more uncertainty one has in the density,
the shorter baseline one has to choose because
the correlation between $\delta$ and $A$
becomes stronger for larger baseline and muon energy.
The works \cite{Freund:2001ui,Pinney:2001xw,huber},
which basically used $\Delta\chi^2$ in (\ref{eqn:chi2}),
agree with each other to certain extent (there are some differences
on the reference values for the oscillation parameters)
while the result by Koike et al.\cite{Koike:2002jf}
is quite different from others.  This discrepancy
is due to the fact that they adopted different
quantity:
\begin{eqnarray}
\Delta{\tilde\chi}^2\equiv
\min_{{\ }_{\overline{\theta_{k\ell}},
\overline{\Delta m^2_{k\ell}},\overline{A}}}
\displaystyle\sum_j {1 \over \sigma^2_j}
\left[
{N_j(\nu_e\rightarrow\nu_\mu) \over
N_j(\bar{\nu}_e\rightarrow\bar{\nu}_\mu)}
-{\bar{N}_j(\nu_e\rightarrow\nu_\mu) \over
\bar{N}_j(\bar{\nu}_e\rightarrow\bar{\nu}_\mu)}
\right]^2,
\label{eqn:chi2kos}
\end{eqnarray}
which turns out to be a combination
in which the correlation between  $\delta$ and $A$
improves for low energy and worsens at high energy.

\begin{figure}
\begin{center}
\vglue -2.5cm \hglue -3cm
\includegraphics[width=6cm]{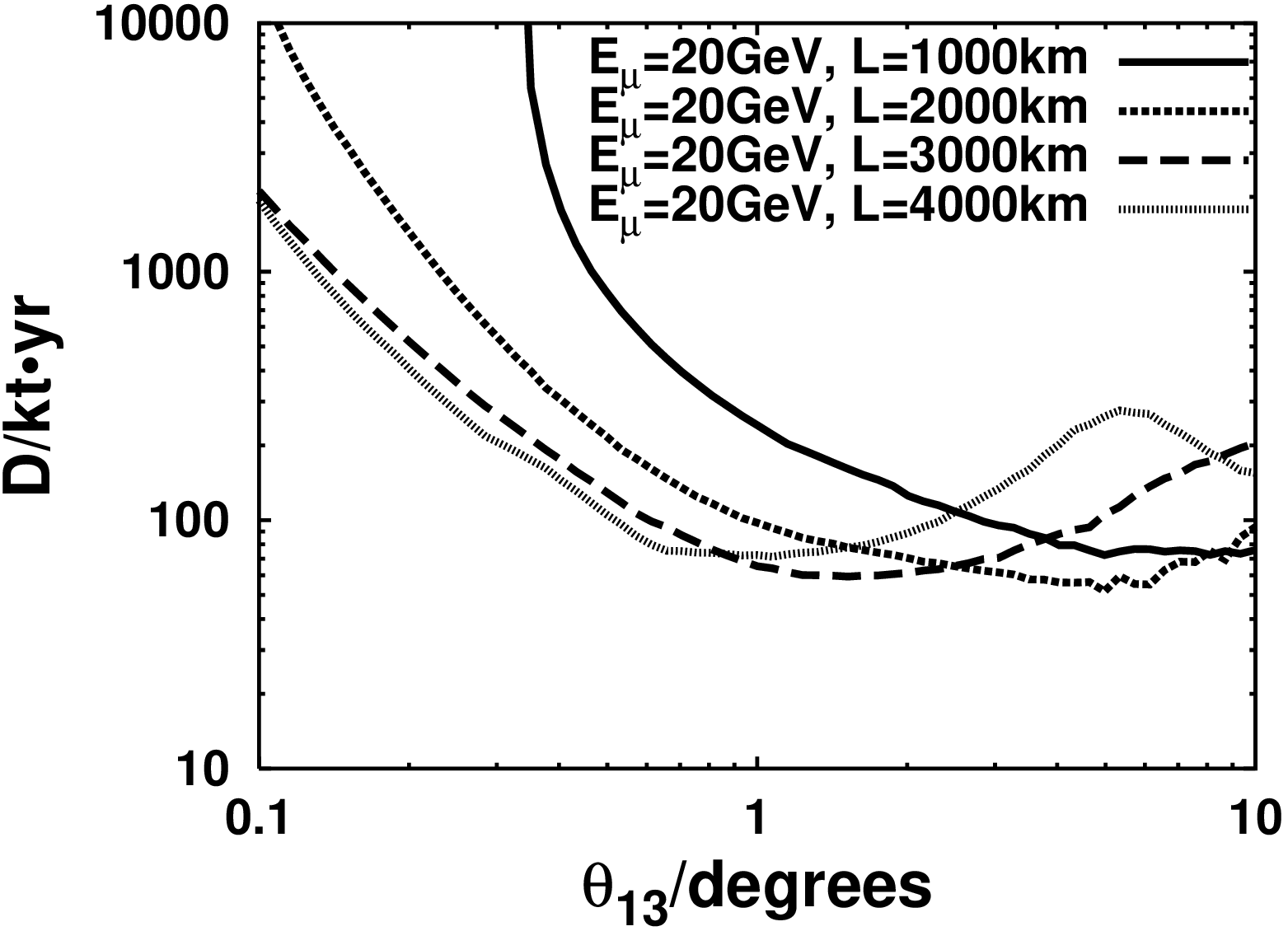}
\vglue -2.5cm \hglue -3cm
\includegraphics[width=6cm]{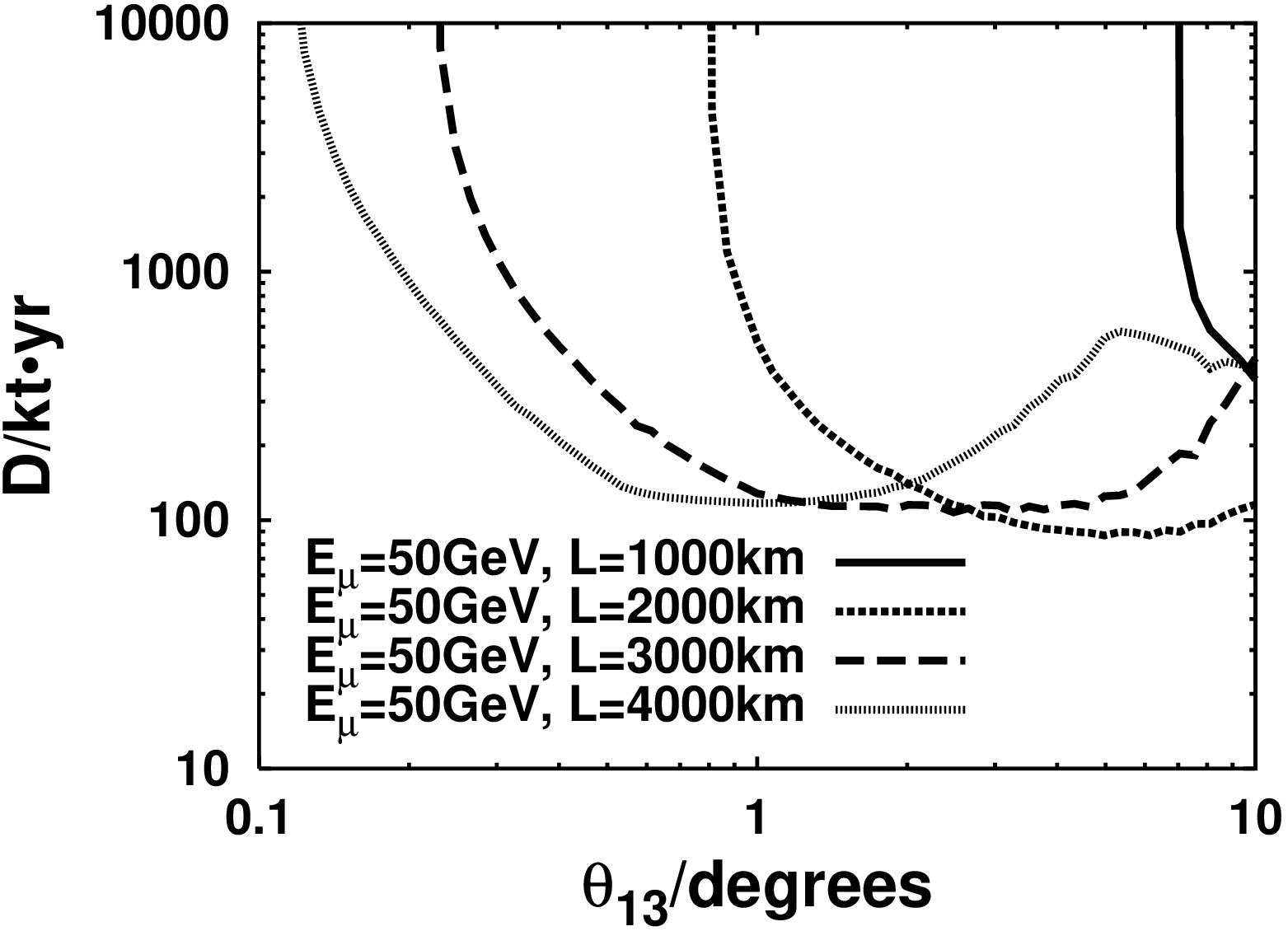}
\vglue 1.0cm
\caption{Data size (kt$\cdot$yr) required to reject a hypothesis of
$\delta=0$ at 3$\sigma$ when the true value is $\delta=\pi/2$,
in the case of a neutrino factory with $10^{21}$ useful muon decays
per year and a background fraction
$f_B=10^{-3}$ \cite{Pinney:2001bj}.}
\label{fig:jp}
\end{center}
\end{figure}

There are slight differences between the results by
the groups \cite{Freund:2001ui,huber} and those by
the other \cite{Pinney:2001xw} and this appears to
come from different statistical treatments.
In the future it should be studied what makes a difference
to get the optimum set $(E_\mu, L)$.
The detector size required to reject a hypothesis "$\delta$=0" as
a function of $\theta_{13}$ is given in
Fig. \ref{fig:jp} (taken from Ref.\cite{Pinney:2001bj}).
Fig. \ref{fig:jp}
shows the sensitivity of neutrino factories to the CP phase.

\subsection{High energy behaviors of $\Delta\chi^2$}
Some people have questioned whether the sensitivity to the CP phase at
a neutrino factory increases infinitely as the muon energy increases,
and Lipari\cite{Lipari:2001ds} concluded that the sensitivity is lost
at high energy.  In the work \cite{Pinney:2001xw} the behaviors
of $\Delta\chi^2$ in (\ref{eqn:chi2}) was studied analytically
for high muon energy and it was shown
after the correlations between $\delta$ and any other oscillation
parameter or $A$ is taken into account that
\begin{eqnarray}
\Delta \chi^2
\propto \left({J \over \sin\delta}\right)^2
{1 \over E_\mu}\left(
\sin\delta+\mbox{\rm const}{\Delta m^2_{32}L \over E_\mu}
\cos\delta\right)^2
\label{eqn:largeemu}
\end{eqnarray}
for large $E_\mu$, where
$J\equiv (c_{13} / 8)\sin2\theta_{12}\sin2\theta_{13}
\sin2\theta_{23}\sin\delta$ stands for the Jarlskog parameter.
The behavior (\ref{eqn:largeemu}) is the same as that for
$\Delta{\tilde\chi}^2$ in (\ref{eqn:chi2kos}) and it is
qualitatively consistent with the claim by
Lipari \cite{Lipari:2001ds}.
It is remarkable that (\ref{eqn:largeemu}) is different from
a naively expected behavior
\begin{eqnarray}
\Delta\chi^2_{\mbox{\rm{\scriptsize  naive}}}\propto E_\mu(\cos\delta-1)^2.
\nonumber
\end{eqnarray}
This is because the correlation between $\delta$ and
other parameters is taken into account.

\section{Parameter degeneracy}

The discussions in the previous sections
have been focused on rejection of
a hypothesis ``$\theta_{13}=0$'' or ``$\delta=0$''.  Once
$\theta_{13}$ or $\delta$ is found to be nonvanishing,
it becomes important
to determine the precise value of $\theta_{13}$ or $\delta$.
Since the work \cite{Burguet-Castell:2001ez}
it has been known that various kinds of parameter degeneracy
exist.  Burguet-Castell et al.\cite{Burguet-Castell:2001ez}
found degeneracy in ($\delta$, $\theta_{13}$),
Minakata and Nunokawa \cite{Minakata:2001qm} found
the one in the sign of $\Delta m^2_{32}$, and
Barger et al.\cite{Barger:2001yr} found the one
in the sign of $\pi/4-\theta_{23}$.
In general, therefore, I have eight-fold degeneracy.
To understand the parameter degeneracy, it is instructive
to consider the appearance probabilities
$P(\nu_\mu\rightarrow\nu_e)$ and
$P(\bar{\nu}_\mu\rightarrow\bar{\nu}_e)$ analytically.
Up to second order in $\alpha$ and $\theta_{13}$ and
assuming constant density of matter, the oscillation
probabilities for $\Delta m^2_{31} > 0$ and $\Delta m^2_{21} > 0$ are
given by \cite{Cervera:2000kp}
\begin{eqnarray}
P\equiv P(\nu_e \to \nu_\mu)
&=&x^2 f^2 + 2 x y f g
(\cos\delta\cos\Delta - \sin\delta\sin\Delta)
+ y^2 g^2\,,\label{eqn:P}\\
\bar{P}\equiv P(\bar\nu_\mu \to \bar\nu_e) &=&
x^2 \bar f^2 + 2 x y \bar f g (\cos\delta\cos\Delta
+ \sin\delta\sin\Delta) + y^2 g^2 \,,\label{eqn:P2}
\end{eqnarray}
respectively, where I follow the notations of
\cite{Barger:2001yr}:
\begin{eqnarray}
x &\equiv& \sin\theta_{23} \sin 2\theta_{13} \,,
\nonumber\\
y &\equiv& \alpha \cos\theta_{23} \sin 2\theta_{12} \,,
\nonumber\\
\left\{ \begin{array}{c}
f\\ \bar f \end{array}\right\}
&\equiv& {\sin((1\mp\hat A)\Delta) \over (1\mp\hat A) }\,,
\nonumber\\
g &\equiv& {\sin(\hat A\Delta) \over \hat A} \,,\nonumber\\
\Delta &\equiv& {|\Delta m_{31}^2| L \over 4E_\nu}
= 1.27 {|\Delta m_{31}^2/{\rm eV^2}| (L/{\rm km}) \over
(E_\nu/{\rm GeV})} \,,
\nonumber\\
\hat A &\equiv& \left|{2AE_\nu \over \Delta m_{31}^2}\right| \,,
\nonumber\\
\alpha &\equiv& \left|{\Delta m^2_{21} \over \Delta m^2_{31}}\right| \,,
\nonumber
\end{eqnarray}
Since (\ref{eqn:P}) and (\ref{eqn:P2}) are quadratic in
$x\equiv \sin\theta_{23} \sin 2\theta_{13}$,
given $P$ and $\bar P$, there are two solutions of (\ref{eqn:P})
with respect to $x$.  In fact it has been known 
\cite{Minakata:2001rj,Kimura:2002hb,Kimura:2002wd,Yokomakura:2002av}
that the solution of (\ref{eqn:P})
constitutes an ellipse in the ($P$, $\bar{P}$) plane
as $\delta$ ranges from 0 to $2\pi$.

\begin{figure}
\vglue 0.7cm 
\hglue -0.4cm
\includegraphics[width=5.5cm]{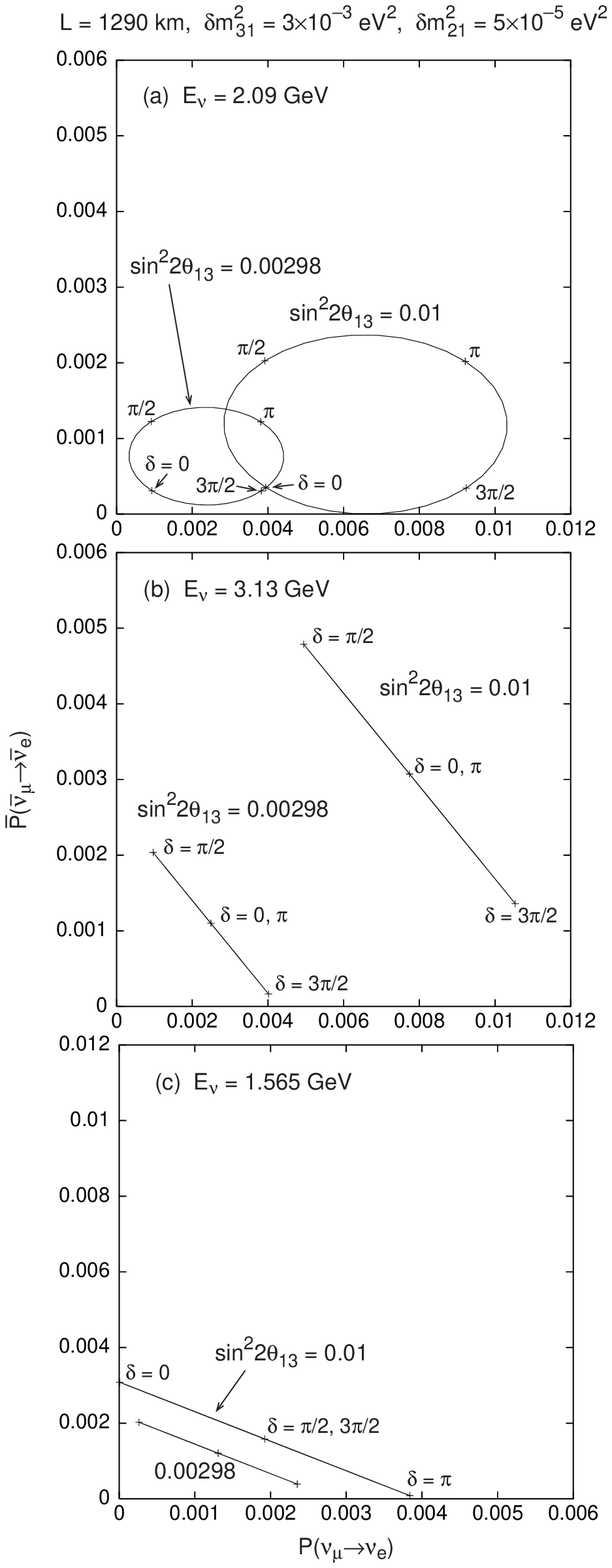}
\vglue -13.9cm 
\hglue 4.8cm
\includegraphics[width=5.5cm]{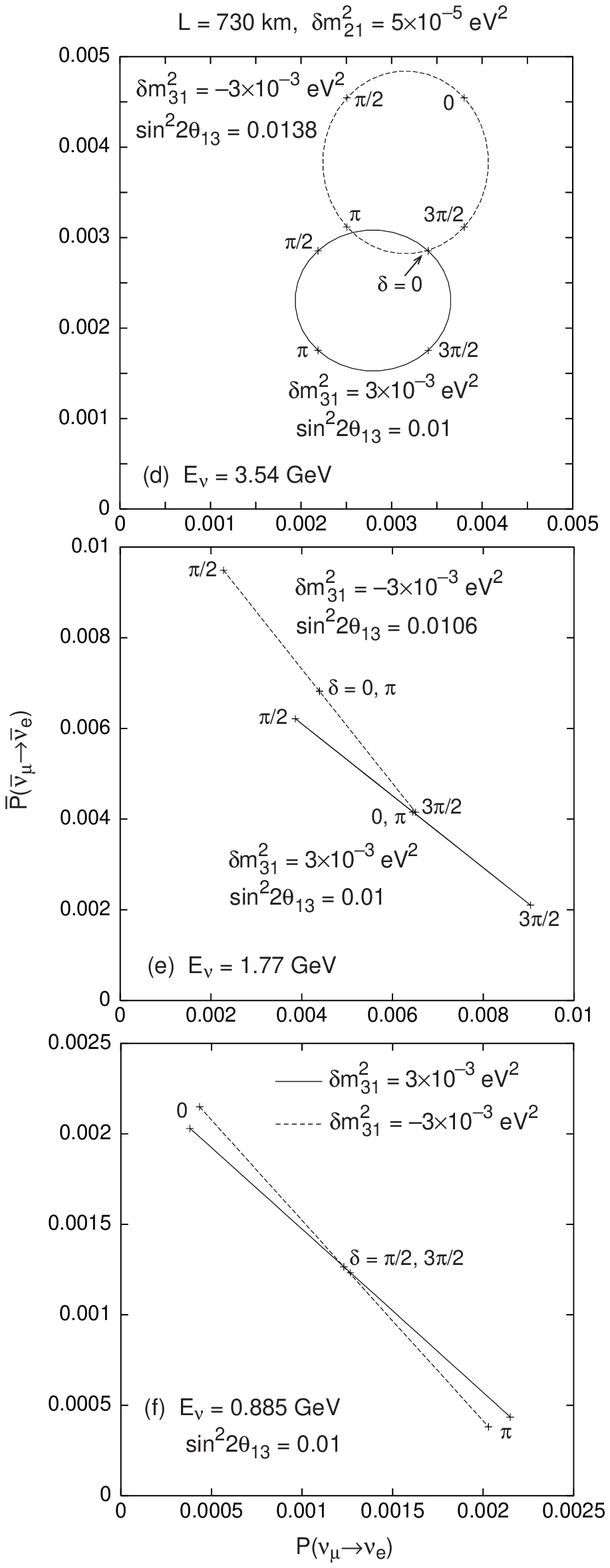}
\vglue -14.2cm 
\hglue 10.3cm
\includegraphics[width=5.5cm]{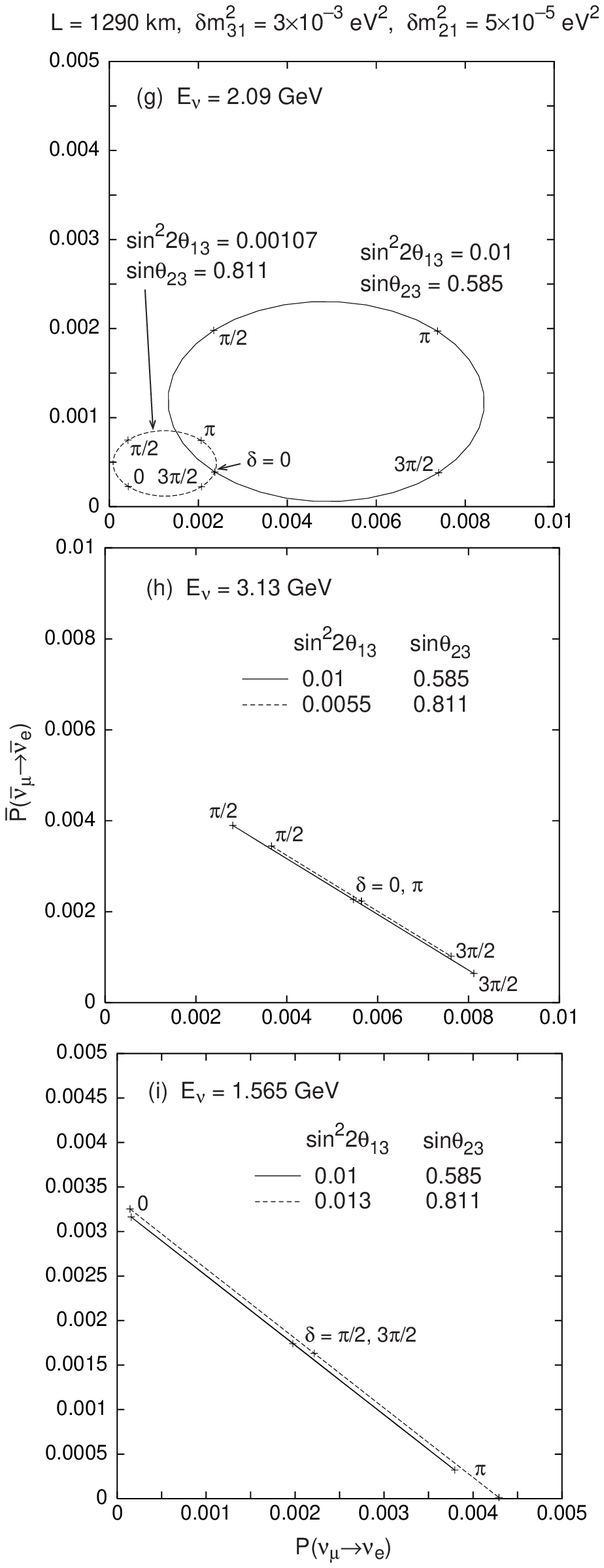}
\hspace*{-20mm}
\caption{The CP trajectory in the ($P$,$\bar{P}$) plane.
(a--c): Orbit ellipses showing $(\delta,\theta_{13})$
ambiguity for $L
= 1290$~km with
(a) $E_\nu = 2.09$~GeV ($\Delta = 3\pi/4$),
(b) $E_\nu = 3.13$~GeV ($\Delta = \pi/2$), and
(c) $E_\nu = 1.565$~GeV ($\Delta = \pi$), for $\sin^22\theta_{13} =
0.01$ and $0.00298$. The other parameters are
$\Delta m^2_{31} = 3\times10^{-3}$~eV$^2$,
$\Delta m^2_{21} = 5\times10^{-5}$~eV$^2$,
$\sin^22\theta_{23} = 1$, and $\sin^22\theta_{12} = 0.8$.
The value of $\delta$ varies around the ellipse.
In (b) and (c) the ellipse collapses to a line and the ambiguity
reduces to a $(\delta, \pi - \delta)$ or $(\delta,2\pi -\delta)$
ambiguity, respectively, and different values of $\theta_{13}$
do not overlap (for the same sgn($\Delta m^2_{13}$)).
(d--f): Sgn($\Delta m^2_{31}$) ambiguity for $L = 730$~km with
(d) $E_\nu = 3.54$~GeV ($\Delta = \pi/4$),
(e) $E_\nu = 1.77$~GeV ($\Delta = \pi/2$), and
(f) $E_\nu = 0.885$~GeV ($\Delta = \pi$).
The other parameters are $\Delta m^2_{21} = 5\times10^{-5}$~eV$^2$,
$\sin^22\theta_{23} = 1$, and $\sin^22\theta_{12} = 0.8$.
(g--i): ($\theta_{23}\,,\pi/2-\theta_{23}$) ambiguity for
$L = 1290$~km with
(g) $E_\nu = 2.09$~GeV ($\Delta = 3\pi/4$),
(h) $E_\nu = 3.13$~GeV ($\Delta = \pi/2$), and
(i) $E_\nu = 1.565$~GeV ($\Delta = \pi$).
The other parameters are $\Delta m^2_{21} = 5\times10^{-5}$~eV$^2$,
and $\sin^22\theta_{12} = 0.8$ \cite{Barger:2001yr}.}
\label{fig:Figbmw}
\end{figure}

\subsection{($\theta_{13}$, $\delta$) ambiguity
\cite{Burguet-Castell:2001ez}}
As the two ellipses indicate in Fig.~\ref{fig:Figbmw}(a),
given $P$ and $\bar P$, there exist two
sets of solution ($\theta_{13}$, $\delta$) and
($\theta^\prime_{13}$, $\delta^\prime$), and
they are given by \cite{Barger:2001yr}
\begin{eqnarray}
x^{\prime 2} - x^2 = {4yg\sin2\Delta
\left[yg\sin2\Delta + xf\sin(\Delta-\delta)
+ x\bar f\sin(\Delta+\delta) \right] \over f^2 + \bar f^2
- 2f\bar f\cos2\Delta} \,,\nonumber
\end{eqnarray}
and $\delta^\prime$ is obtained by
\begin{eqnarray}
x^\prime \cos\delta^\prime &=&
x\cos\delta + {(f+\bar f)(x^2-x^{\prime 2})\over4yg\cos\Delta} \,,
\nonumber\\
x^\prime \sin\delta^\prime &=&
x\sin\delta - {(f-\bar f)(x^2-x^{\prime 2})\over4yg\sin\Delta} \,.
\nonumber
\end{eqnarray}

\subsection{($\Delta m^2_{31}$, $-\Delta m^2_{31}$)
ambiguity \cite{Minakata:2001qm}}
Also for this ambiguity, there are two solutions, as can
be expected from Fig.~\ref{fig:Figbmw}(d).
The general equations for the sgn($\Delta m^2_{31}$) ambiguity
are not easy to work out, so for simplicity let me consider the case
$\Delta = (n-{1\over2})\pi$.
The values
of $(x^\prime,\delta^\prime)$ for $\Delta m^2_{31} < 0$ that give the
same $P$ and $\bar P$ as $(x,\delta)$ for $\Delta m^2_{31} > 0$ are
determined by \cite{Barger:2001yr}
\begin{eqnarray}
x^{\prime2} &=&
{x^2(f^2+\bar f^2-f\bar f)-2yg(f-\bar f)x\sin\delta\sin\Delta
\over f\bar f} \,,\nonumber\\
x^\prime\sin\delta^\prime &=&
x\sin\delta {f^2+\bar f^2-f\bar f\over f\bar f}
- {x^2\over\sin\Delta}{f^2+\bar f^2\over f\bar f}{f-\bar f\over 2yg} \,.
\label{eq:xsind2}
\end{eqnarray}
In particular, if $\sin\delta = 0$ then Eq.~(\ref{eq:xsind2})
gives
\begin{eqnarray}
\sin\delta^\prime =
- x {f^2+\bar f^2\over f\bar f} {f - \bar f\over 2yg\sin\Delta}
\sqrt{f\bar f\over f^2+\bar f^2 - f\bar f} \,,\nonumber
\end{eqnarray}
which indicates that even if $\delta = 0$, there is
some contribution from matter effects which contributes
to CP violation.

\subsection{($\theta_{23}$, $\pi/2-\theta_{23}$)
ambiguity \cite{Barger:2001yr}}
From Fig.~\ref{fig:Figbmw}(g) two solutions are expected.
As with the sgn($\Delta m^2_{31}$) ambiguity, the solutions for the
$\theta_{23}$ ambiguity are complicated in the general case. For the
special case $\Delta = (n-{1\over2})\pi$, I have \cite{Barger:2001yr}
\begin{eqnarray}
\sin^22\theta_{13}^{\prime} &=&
\sin^22\theta_{13} \tan^2\theta_{23}
+ {\alpha^2 g^2 \sin^22\theta_{12} \over f\bar f}(1-\tan^2\theta_{23}) \,,\nonumber\\
\sin2\theta_{13}^\prime \sin\delta^\prime &=&
\sin2\theta_{13} \sin\delta
+ {\alpha g(f-\bar f)\sin2\theta_{12} \over f\bar f}
{\cot2\theta_{23} \over\sin\Delta} \,,\nonumber
\end{eqnarray}
where ($\delta, \theta_{13}$) are the parameters that give a certain
$(P,\bar P)$ for $0 < \theta_{23} < \pi/4$ and ($\delta^\prime,
\theta_{13}^\prime$) are the parameters that give the same $(P,\bar P)$
for $\pi/2 - \theta_{23}$.  Again in this case
$\sin\delta =0$ does not necessarily imply
$\sin\delta^\prime = 0$.

\subsection{Resolution of the degeneracies}
There have been several works to resolve the degeneracies
mainly in the context of conventional beams
\cite{Barger:2001qs,Kajita:2001sb,Donini:2002rm,
Barger:2002rr,Burguet-Castell:2002qx}.

If one tunes the neutrino energy so that 
$|\Delta m_{31}^2| L /4E_\nu = \pi/2$
then the ellipse collapses to a line as depicted in
\ref{fig:Figbmw}(b),(c) and the degeneracy in
($\theta_{13}$, $\delta$) is
lifted \cite{Barger:2001qs,Kajita:2001sb},
assuming that $\theta_{23}$ is close to $\pi/4$.
If one has experiments at different neutrino energies
then the four-fold degeneracy
($\theta_{13}$, $\delta$)$\times$sgn($\Delta m^2_{13}$))
can be removed \cite{Barger:2002rr}.
These scenarios assume approximately monoenergetic
beams in conventional neutrino beams.
In the case of a neutrino factory,
since the neutrino energy spectrum is continuous,
these techniques may not be applicable
unless the energy resolution is good enough
to distinguish the neutrino energies of different bins.

As was discussed in Sect.~\ref{sec:sign},
if one performs a very long baseline experiment with
$L\sim$ 5000km then the sign of $\Delta m^2_{31}$
can be definitely determined, so eventually
the degeneracy in sgn($\Delta m^2_{13}$)) will be resolved.

As for the one in ($\theta_{23}$, $\pi/2-\theta_{23}$),
the most effective way to lift the degeneracy is to
measure the appearance probability $P(\nu_e\rightarrow\nu_\tau)$,
which is proportional to $\cos^2\theta_{23}$ assuming that the
contribution from $\Delta m^2_{21}$ is negligible.
A neutrino factory may be the only experiment which can
resolve this degeneracy.

\section{Summary}
Neutrino factories can establish the existence of
the non-zero CP phase $\delta$ with
the detector size larger than $10^{21}\mu\cdot$100kt$\cdot$yr
for $\sin^22\theta_{13}\gtrsim 10^{-3}$ unless
$|\delta|$ is small.  To determine the precise
value of $\theta_{13}$ and $\delta$, one needs
to resolve the eight-fold degeneracy, and
it will require more than one neutrino energy
and more than one channel (and maybe more than
one baseline) to lift the degeneracy completely.
Comprehensive study on the degeneracy
by taking into account statistical and
systematic errors as well as the error correlations
in the case of a neutrino factory
still needs to be done.

\section*{Acknowledgments}
The author would like to thank H. V. Klapdor-Kleingrothaus,
J. Peltoniemi and other organizers for invitation and hospitality
during the conference.  This
research was supported in part by a Grant-in-Aid for Scientific
Research of the Ministry of Education, Science and Culture,
\#12047222, \#13640295.



\section*{References}
\begin{thebibliography}{99}

\bibitem{Kajita:2001mr}
T.~Kajita and Y.~Totsuka,
Rev.\ Mod.\ Phys.\  {\bf 73}, 85 (2001).

\bibitem{Bahcall:2000kh}
J.~N.~Bahcall,
Phys.\ Rept.\  {\bf 333}, 47 (2000).

\bibitem{Ahmad:2002ka}
Q.~R.~Ahmad {\it et al.}  [SNO Collaboration],
Phys.\ Rev.\ Lett.\  {\bf 89}, 011302 (2002)
[arXiv:nucl-ex/0204009].

\bibitem{Hagiwara:pw}
K.~Hagiwara {\it et al.}  [Particle Data Group Collaboration],
Phys.\ Rev.\ D {\bf 66}, 010001 (2002).

\bibitem{Maki:1962mu}
Z.~Maki, M.~Nakagawa and S.~Sakata,
Prog.\ Theor.\ Phys.\  {\bf 28} (1962) 870.

\bibitem{Pontecorvo:1968fh}
B.~Pontecorvo,
Sov.\ Phys.\ JETP {\bf 26} (1968) 984
[Zh.\ Eksp.\ Teor.\ Fiz.\  {\bf 53} (1968) 1717].

\bibitem{Apollonio:1999ae}
M.~Apollonio {\it et al.}  [CHOOZ Collaboration],
Phys.\ Lett.\ B {\bf 466}, 415 (1999)
[arXiv:hep-ex/9907037].

\bibitem{Cabibbo:1977nk}
N.~Cabibbo,
Phys.\ Lett.\ B {\bf 72}, 333 (1978).

\bibitem{Barger:1980jm}
V.~D.~Barger, K.~Whisnant and R.~J.~Phillips,
Phys.\ Rev.\ Lett.\  {\bf 45}, 2084 (1980).

\bibitem{Pakvasa:1980bz}
S.~Pakvasa,
in {\it C80-07-17.127}
UH-511-410-80
{\it Plenary Session talk given at 20th Int. Conf. on High Energy Physics, Madison, Wis., Jul 17-23, 1980}.

\bibitem{Bilenky:1981hf}
S.~M.~Bilenky and F.~Niedermayer,
Sov.\ J.\ Nucl.\ Phys.\  {\bf 34}, 606 (1981)
[Yad.\ Fiz.\  {\bf 34}, 1091 (1981)].

\bibitem{Itow:2001ee}
Y.~Itow {\it et al.},
arXiv:hep-ex/0106019.

\bibitem{Geer:1997iz}
S.~Geer,
Phys.\ Rev.\ D {\bf 57}, 6989 (1998)
[Erratum-ibid.\ D {\bf 59}, 039903 (1999)]
[arXiv:hep-ph/9712290].

\bibitem{DeRujula:1998hd}
A.~De Rujula, M.~B.~Gavela and P.~Hernandez,
Nucl.\ Phys.\ B {\bf 547}, 21 (1999)
[arXiv:hep-ph/9811390].

\bibitem{Ayres:1999ug}
D.~Ayres {\it et al.}  [Neutrino Factory and Muon Collider Collaboration],
arXiv:physics/9911009.

\bibitem{Albright:2000xi}
C.~Albright {\it et al.},
arXiv:hep-ex/0008064.

\bibitem{Adams:2001tv}
T.~Adams {\it et al.},
in {\it Proc. of the APS/DPF/DPB Summer Study on the Future of Particle Physics (Snowmass 2001) } ed. R.~Davidson and C.~Quigg,
arXiv:hep-ph/0111030.

\bibitem{Hernandez:ki}
P.~Hernandez and O.~Yasuda,
Nucl.\ Instrum.\ Meth.\ A {\bf 485}, 811 (2002).

\bibitem{Yasuda:2001ip}
O.~Yasuda,
arXiv:hep-ph/0111172.

\bibitem{Cervera:2000vy}
A.~Cervera, F.~Dydak and J.~Gomez Cadenas,
Nucl.\ Instrum.\ Meth.\ A {\bf 451}, 123 (2000).

\bibitem{Fukugita:1999as}
M.~Fukugita, G.~C.~Liu and N.~Sugiyama,
Phys.\ Rev.\ Lett.\  {\bf 84}, 1082 (2000)
[arXiv:hep-ph/9908450].

\bibitem{Elgaroy:2002bi}
O.~Elgaroy {\it et al.},
Phys.\ Rev.\ Lett.\  {\bf 89}, 061301 (2002)
[arXiv:astro-ph/0204152].

\bibitem{Klapdor-Kleingrothaus:2001ke}
H.~V.~Klapdor-Kleingrothaus, A.~Dietz, H.~L.~Harney and I.~V.~Krivosheina,
Mod.\ Phys.\ Lett.\ A {\bf 16}, 2409 (2001)
[arXiv:hep-ph/0201231].

\bibitem{Yasuda:1998sf}
O.~Yasuda,
arXiv:hep-ph/9809205.

\bibitem{Mikheev:wj}
S.~P.~Mikheev and A.~Y.~Smirnov,
Nuovo Cim.\ C {\bf 9}, 17 (1986).

\bibitem{Wolfenstein:1977ue}
L.~Wolfenstein,
Phys.\ Rev.\ D {\bf 17}, 2369 (1978).

\bibitem{Yasuda:1999zf}
O.~Yasuda,
arXiv:hep-ph/0005134.

\bibitem{Barger:2000yn}
V.~D.~Barger, S.~Geer, R.~Raja and K.~Whisnant,
Phys.\ Rev.\ D {\bf 62}, 073002 (2000)
[arXiv:hep-ph/0003184].

\bibitem{Barger:2000cp}
V.~D.~Barger, S.~Geer, R.~Raja and K.~Whisnant,
Phys.\ Lett.\ B {\bf 485}, 379 (2000)
[arXiv:hep-ph/0004208].

\bibitem{Cervera:2000kp}
A.~Cervera, A.~Donini, M.~B.~Gavela, J.~J.~Gomez Cadenas, P.~Hernandez, O.~Mena and S.~Rigolin,
Nucl.\ Phys.\ B {\bf 579}, 17 (2000)
[Erratum-ibid.\ B {\bf 593}, 731 (2000)]
[arXiv:hep-ph/0002108].

\bibitem{Kuo:1987km}
T.~K.~Kuo and J.~Pantaleone,
Phys.\ Lett.\ B {\bf 198}, 406 (1987).

\bibitem{Krastev:yu}
P.~I.~Krastev and S.~T.~Petcov,
Phys.\ Lett.\ B {\bf 205}, 84 (1988).

\bibitem{Toshev:vz}
S.~Toshev,
Phys.\ Lett.\ B {\bf 226}, 335 (1989).

\bibitem{Toshev:ku}
S.~Toshev,
Mod.\ Phys.\ Lett.\ A {\bf 6}, 455 (1991).

\bibitem{Arafune:1996bt}
J.~Arafune and J.~Sato,
Phys.\ Rev.\ D {\bf 55}, 1653 (1997)
[arXiv:hep-ph/9607437].

\bibitem{Sato:1997st}
J.~Sato,
Nucl.\ Phys.\ Proc.\ Suppl.\  {\bf 59}, 262 (1997)
[arXiv:hep-ph/9701306].

\bibitem{Koike:1997dh}
M.~Koike and J.~Sato,
arXiv:hep-ph/9707203.

\bibitem{Barger:1999hi}
V.~D.~Barger, Y.~B.~Dai, K.~Whisnant and B.~L.~Young,
Phys.\ Rev.\ D {\bf 59}, 113010 (1999)
[arXiv:hep-ph/9901388].

\bibitem{Koike:1999hf}
M.~Koike and J.~Sato,
Phys.\ Rev.\ D {\bf 61}, 073012 (2000)
[Erratum-ibid.\ D {\bf 62}, 079903 (2000)]
[arXiv:hep-ph/9909469].

\bibitem{Yasuda:1999uv}
O.~Yasuda,
Acta Phys.\ Polon.\ B {\bf 30}, 3089 (1999)
[arXiv:hep-ph/9910428].

\bibitem{Sato:1999wt}
J.~Sato,
Nucl.\ Instrum.\ Meth.\ A {\bf 451}, 36 (2000)
[arXiv:hep-ph/9910442].

\bibitem{Koike:1999tb}
M.~Koike and J.~Sato,
Phys.\ Rev.\ D {\bf 62}, 073006 (2000)
[arXiv:hep-ph/9911258].

\bibitem{Harrison:2000df}
P.~F.~Harrison and W.~G.~Scott,
Phys.\ Lett.\ B {\bf 476} (2000) 349
[hep-ph/9912435]. 

\bibitem{Sato:1999sb}
J.~Sato,
arXiv:hep-ph/0006127.

\bibitem{Yokomakura:2000sv}
H.~Yokomakura, K.~Kimura and A.~Takamura,
Phys.\ Lett.\ B {\bf 496}, 175 (2000)
[arXiv:hep-ph/0009141].

\bibitem{Koike:2001kv}
M.~Koike, T.~Ota and J.~Sato,
arXiv:hep-ph/0103024.

\bibitem{Akhmedov:2001kd}
E.~K.~Akhmedov, P.~Huber, M.~Lindner and T.~Ohlsson,
Nucl.\ Phys.\ B {\bf 608}, 394 (2001)
[arXiv:hep-ph/0105029].

\bibitem{Rubbia:2001pk}
A.~Rubbia,
arXiv:hep-ph/0106088.

\bibitem{Ota:2001cz}
T.~Ota, J.~Sato and Y.~Kuno,
Phys.\ Lett.\ B {\bf 520}, 289 (2001)
[arXiv:hep-ph/0107007].

\bibitem{Guo:2001yt}
W.~l.~Guo and Z.~z.~Xing,
Phys.\ Rev.\ D {\bf 65}, 073020 (2002)
[arXiv:hep-ph/0112121].

\bibitem{Bueno:2001jd}
A.~Bueno, M.~Campanelli, S.~Navas-Concha and A.~Rubbia,
Nucl.\ Phys.\ B {\bf 631}, 239 (2002)
[arXiv:hep-ph/0112297].

\bibitem{Minakata:2002qi}
H.~Minakata, H.~Nunokawa and S.~Parke,
arXiv:hep-ph/0208163.


\bibitem{Naumov:1991rh}
V.~A.~Naumov,
Sov.\ Phys.\ JETP {\bf 74}, 1 (1992)
[Zh.\ Eksp.\ Teor.\ Fiz.\  {\bf 101}, 3 (1992)].

\bibitem{Naumov:1991ju}
V.~A.~Naumov,
Int.\ J.\ Mod.\ Phys.\ D {\bf 1}, 379 (1992).

\bibitem{Geller:2001ix}
R.~J.~Geller and T.~Hara,
arXiv:hep-ph/0111342.

\bibitem{Bueno:2000fg}
A.~Bueno, M.~Campanelli and A.~Rubbia,
Nucl.\ Phys.\ B {\bf 589}, 577 (2000)
[arXiv:hep-ph/0005007].

\bibitem{Koike:2002jf}
M.~Koike, T.~Ota and J.~Sato,
Phys.\ Rev.\ D {\bf 65}, 053015 (2002)
[arXiv:hep-ph/0011387].

\bibitem{Freund:2001ui}
M.~Freund, P.~Huber and M.~Lindner,
Nucl.\ Phys.\ B {\bf 615}, 331 (2001)
[arXiv:hep-ph/0105071].

\bibitem{Pinney:2001xw}
J.~Pinney and O.~Yasuda,
Phys.\ Rev.\ D {\bf 64}, 093008 (2001)
[arXiv:hep-ph/0105087].

\bibitem{Carter:1981tk}
A.~B.~Carter and A.~I.~Sanda,
Phys.\ Rev.\ D {\bf 23}, 1567 (1981).

\bibitem{Bigi:1981qs}
I.~I.~Bigi and A.~I.~Sanda,
Nucl.\ Phys.\ B {\bf 193}, 85 (1981).

\bibitem{Gronau:1990ka}
M.~Gronau and D.~London,
Phys.\ Rev.\ Lett.\  {\bf 65}, 3381 (1990).

\bibitem{Gronau:1991dp}
M.~Gronau and D.~Wyler,
Phys.\ Lett.\ B {\bf 265}, 172 (1991).

\bibitem{huber}
P.~Huber, private communication.

\bibitem{Pinney:2001bj}
J.~Pinney,
[arXiv:hep-ph/0106210].

\bibitem{Lipari:2001ds}
P.~Lipari,
Phys.\ Rev.\ D {\bf 64}, 033002 (2001)
[arXiv:hep-ph/0102046].

\bibitem{Burguet-Castell:2001ez}
J.~Burguet-Castell, M.~B.~Gavela, J.~J.~Gomez-Cadenas, P.~Hernandez and O.~Mena,
Nucl.\ Phys.\ B {\bf 608}, 301 (2001)
[arXiv:hep-ph/0103258].

\bibitem{Minakata:2001qm}
H.~Minakata and H.~Nunokawa,
JHEP {\bf 0110}, 001 (2001)
[arXiv:hep-ph/0108085].

\bibitem{Barger:2001yr}
V.~Barger, D.~Marfatia and K.~Whisnant,
arXiv:hep-ph/0112119.

\bibitem{Minakata:2001rj}
H.~Minakata and H.~Nunokawa,
arXiv:hep-ph/0111130.

\bibitem{Kimura:2002hb}
K.~Kimura, A.~Takamura and H.~Yokomakura,
Phys.\ Lett.\ B {\bf 537}, 86 (2002)
[arXiv:hep-ph/0203099].

\bibitem{Kimura:2002wd}
K.~Kimura, A.~Takamura and H.~Yokomakura,
arXiv:hep-ph/0205295.

\bibitem{Yokomakura:2002av}
H.~Yokomakura, K.~Kimura and A.~Takamura,
arXiv:hep-ph/0207174.

\bibitem{Barger:2001qs}
V.~D.~Barger, D.~Marfatia and K.~Whisnant,
in {\it Proc. of the APS/DPF/DPB Summer Study on the Future of Particle Physics
 (Snowmass 2001) } ed. R.~Davidson and C.~Quigg,
arXiv:hep-ph/0108090.

\bibitem{Kajita:2001sb}
T.~Kajita, H.~Minakata and H.~Nunokawa,
Phys.\ Lett.\ B {\bf 528}, 245 (2002)
[arXiv:hep-ph/0112345].

\bibitem{Donini:2002rm}
A.~Donini, D.~Meloni and P.~Migliozzi,
arXiv:hep-ph/0206034.

\bibitem{Barger:2002rr}
V.~Barger, D.~Marfatia and K.~Whisnant,
arXiv:hep-ph/0206038.

\bibitem{Burguet-Castell:2002qx}
J.~Burguet-Castell, M.~B.~Gavela, J.~J.~Gomez-Cadenas, P.~Hernandez and O.~Mena,
arXiv:hep-ph/0207080.

\end{thebibliography}
\end{document}